\documentclass[prd,aps,10pt,nofootinbib,twocolumn,superscriptaddress,preprintnumbers,balancelastpage,longbibliography,floatfix]{revtex4-2}

\usepackage{afterpage,amsmath,amssymb}
\usepackage{bbold,blindtext,braket,bbm,booktabs}
\usepackage{dcolumn}
\usepackage{fancyhdr,fontawesome}
\usepackage{graphicx}
\graphicspath{{figures/}}
\usepackage{hyperref}
\usepackage{lipsum,listings,longtable}
\usepackage{makecell,mathtools,multirow}
\usepackage{nicefrac}
\usepackage{tabularx}
\usepackage[dvipsnames]{xcolor}
\usepackage{xspace}

\newcommand{\TabTruth}{
\begin{table}[h]
    \centering
    \begin{tabular}{@{}c@{}}
        \begin{tabular}{ccc}
            \multicolumn{3}{c}{\textbf{Diffuse emission}} \\
            \toprule
            Truth & $A$ & $B$ \\
            \midrule
            $S_\text{pib}$ & 11.2 & 10.0 \\
            $\vec\alpha_\text{pib}$ & Model O & Model O \\
            $S_\text{ICS}$ & 6.5 & 6.0 \\
            $\vec\alpha_\text{ICS}$ & Model O & Model O \\
            $S_\text{bub}$ & 1.4 & 0.2 \\
            $S_\text{iso}$ & 0.4 & 0.3 \\
            $S_\text{psc}$ & 2.7 & 0.5 \\
            $S_\text{GCE}$ & 0.6 & 1.0 \\
            $f_\text{bulge}^{\text{pois}}$ & 0.2 & 0.2 \\
            $\vec\alpha_\text{bulge}^{\text{pois}}$ & Col19 & BB \\
            $\gamma^\text{pois}$ & 1.0 & 0.9 \\
            $A_{1,1}$ & 0 & 0 \\
            $A_{1,0}$ & 0 & 0 \\
            $A_{2,2}$ & 0 & 0 \\
            $A_{2,1}$ & 0 & 0 \\
            $A_{2,0}$ & 0 & 0 \\
            \bottomrule
        \end{tabular}
    \end{tabular}
    \quad
    \begin{tabular}{@{}c@{}}
        \setlength{\tabcolsep}{8pt}
        \begin{tabular}{ccc}
            \multicolumn{3}{c}{\textbf{Point sources}} \\
            \toprule
            Truth & $A$ & $B$ \\
            \midrule
            $S_\text{ps}^{\text{GCE}}$ & 1.0 & 1.5 \\
            $n_1^\text{GCE}$ & 5.0 & 5.5 \\
            $n_2^\text{GCE}$ & 1.0 & 1.5 \\
            $n_3^\text{GCE}$ & -5.5 & -5.5 \\
            $S_{b,1}^\text{GCE}$ & 10.0 & 7.6 \\
            $\lambda^\text{GCE}$ & 0.4 & 0.3 \\
            $f_\text{bulge}^{\text{ps}}$ & 0.8 & 0.3 \\
            $\vec\alpha_\text{bulge}^{\text{ps}}$ & Col19 & BBP \\
            $\gamma^\text{ps}$ & 1.2 & 1.2 \\
            $S_\text{ps}^{\text{dsk}}$ & 0.3 & 1.3 \\
            $z_s$ & 0.6 & 0.5 \\
            $C$ & 6.0 & 2.5 \\
            $n_1^\text{dsk}$ & 5.0 & 5.0 \\
            $n_2^\text{dsk}$ & 1.0 & 1.3 \\
            $n_3^\text{dsk}$ & -5.5 & -5.4 \\
            $S_{b,1}^\text{dsk}$ & 15.0 & 11.0 \\
            $\lambda^\text{dsk}$ & 0.7 & 0.7 \\
            \bottomrule
        \end{tabular}
    \end{tabular}
    \caption{Truth values $A$ and $B$ for the two suites of fixed-truth coverage tests. For $\vec\alpha$ values, only the component corresponding to the shown template is 1, while other components are zero.}
    \label{tab:fixed-truths}
\end{table}
}

\newcommand{\TabPrior}{
\begin{table}[h]
    \centering
    \begin{tabular}{@{}c@{}}
        \setlength{\tabcolsep}{8pt}
        \begin{tabular}{cc}
            \multicolumn{2}{c}{\textbf{Diffuse emission}} \\
            \toprule
            Name & Prior \\
            \midrule
            $S_\text{pib}$ & [0.001, 14] \\
            $\vec\alpha_\text{pib}$ & Dirichlet \\
            $S_\text{ICS}$ & [0.001, 14] \\
            $\vec\alpha_\text{ICS}$ & Dirichlet \\
            $S_\text{bub}$ & [0.001, 5] \\
            $S_\text{iso}$ & [0.001, 5] \\
            $S_\text{psc}$ & [0.001, 5] \\
            $S_\text{GCE}$ & [$10^{-5}$, 4] \\
            $f_\text{bulge}^{\text{pois}}$ & [0, 1] \\
            $\vec\alpha_\text{bulge}^{\text{pois}}$ & Dirichlet \\
            $\gamma^\text{pois}$ & [0.2, 2] \\
            $A_{1, 1}$ & [-0.05, 0.05] \\
            $A_{1, 0}$ & [-0.05, 0.05] \\
            $A_{2, 2}$ & [-0.05, 0.05] \\
            $A_{2, 1}$ & [-0.05, 0.05] \\
            $A_{2, 0}$ & [-0.05, 0.05] \\
            \bottomrule
        \end{tabular}
    \end{tabular}
    \quad
    \begin{tabular}{@{}c@{}}
        \setlength{\tabcolsep}{8pt}
        \begin{tabular}{cc}
            \multicolumn{2}{c}{\textbf{Point sources}} \\
            \toprule
            Name & Prior \\
            \midrule
            $S_\text{ps}^{\text{GCE}}$ & [$10^{-5}$, 4] \\
            $n_1^\text{GCE}$ & [2.1, 8] \\
            $n_2^\text{GCE}$ & [0.5, 2] \\
            $n_3^\text{GCE}$ & [-8, -1] \\
            $S_{b,1}^\text{GCE}$ & [5, 40] \\
            $\lambda^\text{GCE}$ & [0.1, 0.95] \\
            $f_\text{bulge}^{\text{ps}}$ & [0, 1] \\
            $\vec\alpha_\text{bulge}^{\text{ps}}$ & Dirichlet \\
            $\gamma^\text{ps}$ & [0.2, 2] \\
            $S_\text{ps}^{\text{dsk}}$ & [$10^{-5}$, 4] \\
            $z_s$ & [0.1, 2.5] \\
            $C$ & [0.05, 8] \\
            $n_1^\text{dsk}$ & [2.1, 8] \\
            $n_2^\text{dsk}$ & [0.5, 2] \\
            $n_3^\text{dsk}$ & [-8, -1] \\
            $S_{b,1}^\text{dsk}$ & [5, 40] \\
            $\lambda^\text{dsk}$ & [0.1, 0.95] \\
            \bottomrule
        \end{tabular}
    \end{tabular}
    \caption{{Priors for fiducial model} used in \textit{Fermi} fits and simulation tests. See texts in Sec.~\ref{sec:results} for a detailed description of the parameters.}
    \label{tab:prior}
\end{table}
}

\newcommand{\TabPriorRestricted}{
\begin{table}[h]
    \centering
    \begin{tabular}{@{}c@{}}
        \setlength{\tabcolsep}{8pt}
        \begin{tabular}{cc}
            \multicolumn{2}{c}{\textbf{Diffuse emission}} \\
            \toprule
            Name & Prior \\
            \midrule
            $S_\text{pib}$ & [0.001, 12] \\
            $\vec\alpha_\text{pib}$ & Dirichlet \\
            $S_\text{ICS}$ & [0.001, 7] \\
            $\vec\alpha_\text{ICS}$ & Dirichlet \\
            $S_\text{bub}$ & [0.001, 2] \\
            $S_\text{iso}$ & [0.001, 1] \\
            $S_\text{psc}$ & [0.001, 4] \\
            $S_\text{GCE}$ & [$10^{-5}$, 2] \\
            $f_\text{bulge}^{\text{pois}}$ & [0, 1] \\
            $\vec\alpha_\text{bulge}^{\text{pois}}$ & Dirichlet \\
            $\gamma^\text{pois}$ & [0.2, 2] \\
            $A_{1, 1}$ & [-0.05, 0.05] \\
            $A_{1, 0}$ & [-0.05, 0.05] \\
            $A_{2, 2}$ & [-0.05, 0.05] \\
            $A_{2, 1}$ & [-0.05, 0.05] \\
            $A_{2, 0}$ & [-0.05, 0.05] \\
            \bottomrule
        \end{tabular}
    \end{tabular}
    \quad
    \begin{tabular}{@{}c@{}}
        \setlength{\tabcolsep}{8pt}
        \begin{tabular}{cc}
            \multicolumn{2}{c}{\textbf{Point sources}} \\
            \toprule
            Name & Prior \\
            \midrule
            $S_\text{ps}^{\text{GCE}}$ & [$10^{-5}$, 2] \\
            $n_1^\text{GCE}$ & [2.1, 8] \\
            $n_2^\text{GCE}$ & [0.5, 2] \\
            $n_3^\text{GCE}$ & [-8, -1] \\
            $S_{b,1}^\text{GCE}$ & [5, 40] \\
            $\lambda^\text{GCE}$ & [0.1, 0.95] \\
            $f_\text{bulge}^{\text{ps}}$ & [0, 1] \\
            $\vec\alpha_\text{bulge}^{\text{ps}}$ & Dirichlet \\
            $\gamma^\text{ps}$ & [0.2, 2] \\
            $S_\text{ps}^{\text{dsk}}$ & [$10^{-5}$, 2.5] \\
            $z_s$ & [0.1, 2.5] \\
            $C$ & [0.05, 8] \\
            $n_1^\text{dsk}$ & [2.1, 8] \\
            $n_2^\text{dsk}$ & [0.5, 2] \\
            $n_3^\text{dsk}$ & [-8, -1] \\
            $S_{b,1}^\text{dsk}$ & [5, 40] \\
            $\lambda^\text{dsk}$ & [0.1, 0.95] \\
            \bottomrule
        \end{tabular}
    \end{tabular}
    \caption{Restricted priors for SVI marginal calibration test. See text for differences to the fiducial prior.}
    \label{tab:prior-restricted}
\end{table}
}

\lstdefinestyle{python}{
  language=Python,
  basicstyle=\small\ttfamily,
  keywordstyle=\color{MidnightBlue}\bfseries,
  stringstyle=\color{BrickRed},
  commentstyle=\color{OliveGreen}\itshape,
  showstringspaces=false,
  breaklines=true,
  frame=single,
  rulecolor=\color{black!30},
  backgroundcolor=\color{black!3},
  xleftmargin=2pt,
  xrightmargin=2pt,
  aboveskip=6pt,
  belowskip=6pt,
  morekeywords={self,as},
  columns=flexible,
}

\colorlet{linkcolor}{BrickRed}
\hypersetup{colorlinks=true,
linkcolor=linkcolor,
citecolor=linkcolor,
urlcolor=linkcolor,
,linktocpage=true
,pdfproducer=medialab}

\newcommand{\Fermi}{\emph{Fermi}\xspace}

\definecolor{deepgreen}{rgb}{0.2,0.8,0.2}

\definecolor{yscolor}{rgb}{0.0, 0.6, 0.0}

\begin{document}

\preprint{\hfill MIT-CTP/6025}

\title{High-dimensional inference for the $\gamma$-ray sky with differentiable programming}

\author{Siddharth Mishra-Sharma}
\email{smsharma@mit.edu}
\thanks{ORCID: \href{https://orcid.org/0000-0001-9088-7845}{0000-0001-9088-7845}}
\affiliation{Faculty of Computing and Data Sciences, Boston University, Boston, MA 02215, USA}
\affiliation{The NSF AI Institute for Artificial Intelligence and Fundamental Interactions}
\affiliation{Center for Theoretical Physics -- a Leinweber Institute, Massachusetts Institute of Technology, Cambridge, MA 02139, USA}
\affiliation{Department of Physics, Harvard University, Cambridge, MA 02138, USA}

\author{Tracy R.~Slatyer}
\email{tslatyer@mit.edu}
\thanks{ORCID: \href{https://orcid.org/0000-0001-9699-9047}{0000-0001-9699-9047}}
\affiliation{The NSF AI Institute for Artificial Intelligence and Fundamental Interactions}
\affiliation{Center for Theoretical Physics, Massachusetts Institute of Technology, Cambridge, MA 02139, USA}

\author{Yitian Sun}
\email{yitian.sun@mcgill.ca}
\thanks{ORCID: \href{https://orcid.org/0000-0002-4697-8384}{0000-0002-4697-8384}}
\affiliation{Trottier Space Institute \& Department of Physics, McGill University, Montreal, QC H3A 2T8, Canada}

\author{Yuqing Wu}
\email{yw2549@cornell.edu}
\thanks{ORCID: \href{https://orcid.org/0009-0007-2571-7103}{0009-0007-2571-7103}}
\affiliation{Department of Physics, Cornell University, Ithaca, NY 14853, USA}

\date{\today}

\begin{abstract}
    We motivate the use of differentiable probabilistic programming techniques in order to account for the large model-space inherent to astrophysical $\gamma$-ray analyses. Targeting the longstanding Galactic Center $\gamma$-ray Excess (GCE) puzzle, we construct differentiable forward model and likelihood that make liberal use of GPU acceleration and vectorization in order to simultaneously account for a continuum of possible spatial morphologies consistent with the GCE emission in a fully probabilistic manner. Our setup allows for efficient inference over the large model space using variational methods. Beyond application to $\gamma$-ray data, a goal of this work is to showcase how differentiable probabilistic programming can be used as a tool to enable flexible analyses of astrophysical datasets.
\end{abstract}

\maketitle

\tableofcontents

\section{Introduction} 
\label{sec:intro}

Analysis of $\gamma$-ray data towards the Galactic Center involves a subtle interplay between observations and theoretical models. The dominant background arises from cosmic rays interacting with the interstellar gas and radiation field, and is in principle quite well understood physically. However, the full phase space distribution of the inputs (i.e.~the cosmic rays, gas, and radiation) is quite challenging to model and constrain. The complexity of cosmic-ray transport processes in the region, combined with the fact that we only have access to a two-dimensional projection of the inherently three-dimensional Galactic gamma-ray emission, means that we typically cannot make precise predictions about emission even from known sources. When we seek to introduce additional source terms to explain anomalies, that introduces additional degrees of freedom. Consequently, scientific conclusions drawn from $\gamma$-ray analyses can depend strongly on specific assumptions made about the modeling of the signal and background, as well as ad hoc choices involved in the data analysis pipeline.

The $\gamma$-ray Galactic Center Excess (GCE), first identified over a decade ago~\cite{Goodenough:2009gk} using public data from the \Fermi-LAT telescope~\cite{Atwood:2009ez}, is an excess of photons in the Inner Galaxy whose physical origin is currently unknown. Although the GCE has properties broadly compatible with those expected from annihilating dark matter (DM)~\cite{Goodenough:2009gk,Hooper:2010mq,Daylan:2014rsa,Calore:2014xka,Fermi-LAT:2015sau,DiMauro:2021raz,Cholis:2021rpp,McDermott:2022zmq,Zhong:2024vyi,Muru:2025vpz}, competing explanations in terms of a population of unresolved astrophysical point sources (PSs), in particular millisecond pulsars, remain viable~\cite{Abazajian:2014fta,Abazajian:2010zy} and are favored by the data in some analyses~\cite{Macias:2016nev,Bartels:2017vsx, Macias:2019omb, Song:2024iup}. Significant effort has gone into characterizing the GCE and understanding its origin, including through inferring the properties of unresolved PSs in the Inner Galaxy region. A particularly successful method in this direction, Non-Poissonian Template Fitting (NPTF), aims to characterize different populations of unresolved PSs using their photon statistics, via a likelihood-based analysis~\cite{Lee:2014mza,Lee:2015fea}. In the context of the GCE, initial applications of this method showed overwhelming support for the hypothesis of the GCE being sourced by an unresolved population of PSs spatially correlated with the observed GCE morphology~\cite{Lee:2015fea}. However, more recent studies showed the method to be susceptible to systematic biases~\cite{Leane:2019xiy,Leane:2020nmi,Leane:2020pfc,Chang:2019ars} partly stemming from degeneracies between different emission components in the region (e.g., those spatially correlated with the GCE, centrally-concentrated stellar populations, and the Galactic disk), with results depending strongly on specific assumptions made about the spatial morphologies of components. All NPTF analyses to date have assumed inflexible, spatially rigid templates, preventing a rigorous expression and exploration of this degeneracy.
\vspace{0.05in}

In this work, we use differentiable probabilistic programming as a means to alleviate this shortcoming of traditional likelihood-based fitting techniques for analyzing $\gamma$-ray data. Using GPU-native acceleration and vectorization, we construct a probabilistic forward model for $\gamma$-ray emission in the Galactic Center region that can generate high-resolution spatial templates for various emission components, including those associated with the GCE signal, on the fly during inference. This framework enables analyses that can be more resilient to specific modeling choices by including a range of physically-reasonable variations on different emission components directly at the level of the forward model. An early version of this pipeline has already been applied to study the GCE morphology by modeling the signal with a Gaussian process \cite{Ramirez:2024oiw}, albeit without implementation of unresolved PS populations. In the broader context of astrophysical data analysis, this work aims to motivate the use of differentiable probabilistic programming as a way to enable high-dimensional Bayesian inference for complex descriptions of diverse astrophysical data.

We describe our methodology, including the construction of the model and the implementation of our inference pipeline, in Sec.~\ref{sec:methodology}. We apply the pipeline to the real {\it Fermi}-LAT data in Sec.~\ref{sec:results}, and then run validation tests on simulated data (with truth values motivated by the fit to the real data) in Sec.~\ref{sec:validation}. We present our conclusions in Sec.~\ref{sec:conclusion}. 

\section{Methodology} 
\label{sec:methodology}

In this section we describe our forward model and likelihood (which exactly models the diffuse emission but only approximately models the unresolved point source populations), its implementation as a differentiable probabilistic program, as well as the inference strategies used to perform Bayesian posterior inference.

\subsection{The model and likelihood}
    
In traditional spatial template fitting, photon counts $x^p$ in different pixels $p$ are modeled as Poisson random variables, with mean given by a linear combination of templates $T_i^p$, indexed by $i$; $x^p \sim \operatorname{Pois}\left(x^p \mid \sum_i A_i T_i^p\right)$. $A_i$ are the template normalizations and represent the parameters of interest. The full map-level likelihood factorizes into a product of pixel-wise likelihoods, $p(x \mid \{A_i\})= \prod_p p(x_p \mid \{A_i\})$.

In the presence of populations of unresolved PSs, the likelihood can be augmented to take into account the probability that a given pixel contains a PS from a given population, and the probability of that PS emitting a certain number of contributing photons. Point source populations can be independently characterized by a spatial distribution, specified by a template $T_{\mathrm{PS}}^p$, and a source count function (SCF) ${\mathrm{d} N} / {\mathrm{d} S}$, which factorizes to describe the distribution of photon counts $S$ from individual PSs, $p(S | \theta_\mathrm{PS})$, and their mean abundance, $\bar n_\mathrm{PS}$, as ${\mathrm{d} N} / {\mathrm{d} S}=\bar n_\mathrm{PS} \, p(S \mid \theta_{\mathrm{PS}})$. 

There are several levels at which one could model ${\mathrm{d} N} / {\mathrm{d} S}$. Given an intrinsic luminosity function for the source population and a spatial distribution, one could integrate along the line of sight to determine the (expected) number of sources contributing a certain flux incident on the detector. Given the large uncertainty in the underlying luminosity functions, we do not take this approach in this work; instead, consistent with previous NPTF studies, we simply model the flux distribution with a parametric function. The conversion between the flux incident on the detector and the number of observed counts requires the detector exposure, which may differ across the relevant area of the sky. Following \cite{Mishra-Sharma:2016gis}, we define our parameters to specify  ${\mathrm{d} N} / {\mathrm{d} S}$ at the mean exposure, but then rescale the parameters according to the ratio of the position-dependent exposure and the mean exposure, such that the ${\mathrm{d} N} / {\mathrm{d} S}$ function used in the inference pipeline corresponds to a fixed source distribution with respect to photon flux incident on the detector (i.e. the observed count distribution varies as one would expect from the exposure).

Unfortunately, the inclusion of PS populations renders the exact likelihood intractable, since computing it would involve marginalizing over positions, photon counts, and the number of PSs from each population. The Non-Poissonian Template Fitting (NPTF) method~\cite{Lee:2014mza,Lee:2015fea} circumvents this issue by computing the pixel-wise likelihood accounting for the probability of finding a given number of sources in that pixel, but without seeking to include the positions of individual sources as latent variables; the full likelihood is approximated in a factorized form, $p\left(x \mid \bar n_\mathrm{PS}, \theta_{\mathrm{PS}}\right)=\prod_p p\left(x^p \mid \bar n_\mathrm{PS}, \theta_{\mathrm{PS}}\right)$ (see \citet{Mishra-Sharma:2016gis} for further details on the method). We implement a differentiable version of the NPTF likelihood using \texttt{Jax}~\cite{jax2018github}, making liberal use of vectorization and automatic batching to efficiently compute the likelihood for multiple PS populations.

\subsection{Flexible specification of point source populations}

Previous analyses based on the NPTF framework assumed rigidly specified spatial templates for the distribution of PS populations; this is because generating new morphologies adds considerable overhead in the likelihood calculation, rendering a traditional Monte Carlo sampling-based analysis difficult. In particular, given a 3D morphology for the PS template, computing the template expectation in a given pixel $p$ involves performing a line-of-sight integral in the pixel direction. In \texttt{Jax}, the density over the line of sight as well as across the pixel index can be efficiently vectorized, and the computation is just-in-time compiled for efficient execution on a GPU.

This allows us to include parametric variations on different spatial templates that have been previously included in NPTF analyses, including templates for PSs correlated with the Galactic disk and the GCE, directly as part of the inference pipeline. This flexibility also applies to the simpler templates associated with diffuse emission, with purely Poissonian statistics.

We model a population of PSs spatially correlated with the Galactic disk, using a cylindrically symmetric, doubly exponential profile~\cite{bartels2018bayesian,lorimer2006parkes}, $\rho_{\text {disk }}(R, z) \propto {R}^B\exp \left[-C\left({R-R_{\odot}}\right)/{R_\odot}\right] \exp \left(-{|z|}/{z_s}\right)$ with the parameter $B$, $C$, and the scale height $z_s$ characterizing the disk morphology to be inferred during the analysis. $R_{\odot}$ is the distance of the Sun from the Galactic Center.

For the GCE, the possibility of gamma-ray emission correlated with the stellar density in the Galactic bulge has been extensively tested in the literature~\cite{Macias:2016nev,Macias:2019omb,Ploeg:2020jeh}, with different analyses finding divergent results for whether this scenario is preferred compared to a spherically symmetric profile. We allow for this possibility in our analysis by including point source populations tracing multiple publicly-available spatial templates following stellar bulge populations~\cite{Macias:2016nev,coleman2020maximum,Macias:2019omb,mcdermott2023morphology}. We also model a PS population with a generalized Navarro-Frenk-White (NFW) profile, motivated by the DM expectation:
$\rho_{\mathrm{NFW}}(r) \propto\left({r}/{r_s}\right)^{-\gamma}\left(1+{r}/{r_s}\right)^{-3+\gamma}$ with the inner slope $\gamma$ a free parameter to be inferred during the analysis, and $r_s$ a fixed scale radius. The general expectation is that DM-sourced emission should be smooth and diffuse (although see \cite{Agrawal:2017pnb}), but we wish to ensure that any preference for PS-like vs smooth/diffuse emission is independent from a preference for a specific morphology, and so we test a range of morphologies for both PS-like and smooth components. This choice also allows direct comparison of our pipeline with previous NPTF analyses of the GCE, which typically used a NFW-motivated template for the PS population.

\begin{figure}[ht!]
\centering
\includegraphics[width=0.48\textwidth]{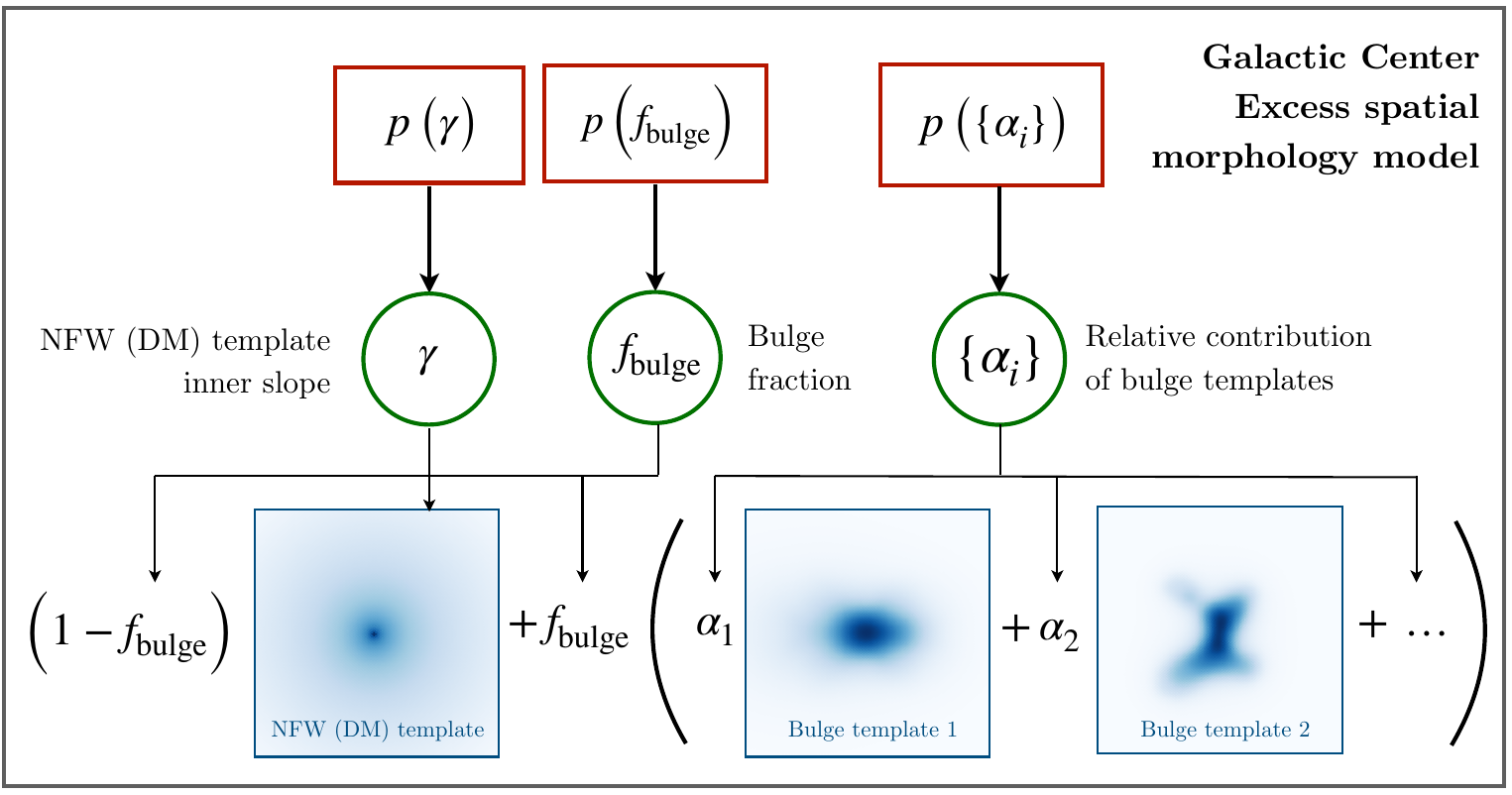}
\caption{Graphical model of the spatial morphology of the Galactic Center Excess, as specified for both the PS as well as smooth emission components.}
\label{fig:pgm}
\end{figure}

For both the PS as well as smooth (Poissonian) components describing the GCE, we model the spatial distribution as a hybrid template allowing for varying fractions of DM-like and bulge-like morphologies, $T_{\mathrm{GCE}}^p=(1 - f_\mathrm{bulge}) T_{\mathrm{NFW}}^p+f_\mathrm{bulge} T_{\mathrm{bulge}}^p$. The bulge template itself is modeled as a linear combination of publicly available bulge morphologies, $T_{\mathrm{bulge}}^p=\sum_i \alpha_i T_{\mathrm{bulge}, i}^p$ with $\alpha_i$ modeled with a symmetric Dirichlet prior, enforcing $\sum_i \alpha_i = 1$. The templates are appropriately normalized such that an un-scaled template would produce one count per pixel within the region of interest (ROI), allowing for a physically-meaningful interpretation of the bulge fraction $f_\mathrm{bulge}$ parameter as the relative contribution to the GCE PS or smooth component from the composite bulge component. 
A graphical representation of the model of the spatial morphology of the GCE, as specified for both the PS as well as smooth components, is shown in Fig.~\ref{fig:pgm}. The complete specification of all model parameters and their priors for our fiducial analysis is given in Sec.~\ref{sec:results}.

\subsection{Parameter posterior inference}

Since the entire pipeline from template generation to likelihood evaluation is end-to-end differentiable, we can evaluate gradients of the log-likelihood function with respect to the parameters of interest $\theta$ characterizing the smooth as well as PS components, $\nabla_\theta\log p(x\mid\theta)$, with minimal overhead. This opens up the possibility of using highly efficient gradient-based posterior inference techniques like Stochastic Variational Inference (SVI)~\cite{2012arXiv1206.7051H} and Hamiltonian Monte Carlo (HMC)~\cite{2011arXiv1111.4246H}. 
Unlike traditional sampling techniques previously employed in the context of \Fermi GCE analyses, these methods can easily scale up to high-dimensional parameter spaces. In particular, variational methods turn posterior inference into an optimization problem by fitting for a flexible functional ansatz on the posterior density and can easily deal with hundreds of parameters. This is in contrast to current public implementations of NPTF, which cannot efficiently include real-time generated templates at inference and are restricted to a modest number of parameters with traditional Monte Carlo sampling techniques. 

The constructed likelihood is wrapped using the probabilistic programming framework \texttt{NumPyro}~\cite{2019arXiv191211554P}, which allows for flexible specification of complex forward models as well as performing inference on them using gradient-based techniques.

In SVI an approximation for the posterior distribution, parameterized by $\varphi$ and denoted $q_\varphi(\theta)$, is obtained by minimizing the reverse KL-divergence between the true and variational posterior distributions, $D_\mathrm{KL}\left(q_\varphi\mid\mid p(\theta\mid x)\right)$. This is done by maximizing the model log-evidence $\log p(x)$ or, in practice using the tractable evidence lower bound (ELBO), 
$\operatorname{ELBO} \equiv \mathbb E_{\vartheta\sim q_\varphi(\theta)}\left[\log p(x, \theta) - \log q_\varphi(\theta)\right]$,
as the optimization objective since $\log p(x) - \operatorname{ELBO} = D_\mathrm{KL}\left(q_\varphi\mid\mid p(\theta\mid x)\right)$. The expectation in the ELBO is taken through Monte Carlo sampling from the variational distribution at each optimization step.

Since we seek the capability to model arbitrarily flexible non-Gaussian posterior distributions, the variational family $q_\varphi$ is modeled using an inverse autoregressive normalizing flow (IAF)~\cite{10.5555/3157382.3157627} consisting of 5 flow transformations modeled using masked autoencoders consisting of two 128-dimensional hidden layers with tanh nonlinearities.
The variational parameters -- the parameters of the normalizing flow transformation -- are optimized using \texttt{Optax}~\citep{optax2020github} with the Adam optimizer~\cite{kingma2014adam,loshchilov2017decoupled}. With an initial learning rate of $1\sim3\times10^{-4}$, we found that 5000-10000 steps is sufficient for convergence. 

For HMC, we use the No-U-Turn Sampler (NUTS)~\cite{hoffman2014no} as implemented in \texttt{NumPyro}. Our fiducial configuration uses an initial step size of $0.05$, with $1000$ warmup steps per chain. We caution that significant modifications to the fiducial model may require changes to these specific parameters of the samplers.

\subsection{Public implementation of the pipeline}

The code used in this work is publicly available at \url{https://github.com/smsharma/fermi-prob-prog}. Fitting the fiducial model to \textit{Fermi} data requires only a few lines:
\begin{lstlisting}[style=python]
m = NPModel()  # load data, templates, PSF, masks
# fit with SVI
m.fit_svi(data=m.fermi_data, lr=3e-4, n_steps=5000)
# or NUTS HMC
m.run_nuts(data=m.fermi_data, num_chains=4, num_samples=5000)
\end{lstlisting}
The probabilistic model, encoded in \texttt{NPModel.model()}, is straightforward to extend. For instance, an additional smooth Poissonian template can be included by sampling its normalization and adding it to the total Poissonian expectation value:
\begin{lstlisting}[style=python]
def model(self, data=None):
    # ...
    S_new = numpyro.sample("S_new",
                     dist.Uniform(1e-5, 5.))
    mu += S_new * temp_new # simply add to total
    # ...
\end{lstlisting}
Meanwhile, adding a new PS population requires appending its spatial template and SCF parameters to existing structures:
\begin{lstlisting}[style=python]
def model(self, data=None):
    # ...
    Sps = numpyro.sample("Sps_new",
                     dist.Uniform(1e-5, 8.))
    n1  = numpyro.sample("n1_new",
                     dist.Uniform(2.1, 8.))
    # ... and n2, n3, sb1, lams analogously
    A = ... # compute SCF norm A from Sps
    # append to existing structure
    npt_compressed.append(temp_new_ps)
    theta.append([A, n1, n2, n3, sb1, sb1*lams])
    # ...
\end{lstlisting}
Simulated data can be generated by passing the desired ground truth values to \texttt{NPModel.simulate()} as
\begin{lstlisting}[style=python]
counts_map = m.simulate({"S_pib": 10., ...})
\end{lstlisting}
Complete working examples are provided in the repository. In addition to this repository, an energy-dependent, purely Poissonian version of the pipeline---applicable to analyses that do not require non-Poissonian PS characterization---is separately available at \url{https://github.com/yitiansun/gce-prob-prog} and has been modified for use in Ref.~\cite{Ramirez:2024oiw}.

\section{Results on \Fermi data} 
\label{sec:results}

\subsection{Dataset and templates}
\label{subsec:dataset-templates}

We apply our pipeline to 573 weeks of \Fermi-LAT data in the 2--20\,GeV energy range. Following previous NPTF analyses (in particular Ref.~\cite{Mishra-Sharma:2016gis} and the public \texttt{NPTFit} code package presented in that work), we restrict to the \texttt{ULTRACLEANVETO} event class (corresponding to the most stringent cosmic ray rejection) and to the top quartile of events graded by angular resolution. We impose a standard zenith angle cut of 90$^\circ$, and the standard quality cuts \texttt{DATA\_QUAL==1 \&\& LAT\_CONFIG==1}. For purposes of the NPTF likelihood computation, in the analyses presented in this work, we divide the exposure map into 7 regions following the method in NPTFit \cite{Mishra-Sharma:2016gis}. More generally, the number of regions is an adjustable parameter in our pipeline; our default choice of 7 was based on early tests indicating the results were stable for 5+ exposure regions.

Smooth templates corresponding to isotropically-uniform emission, emission from (resolved) PSs in the 3FGL catalog~\cite{Fermi-LAT:2015bhf}, and the \Fermi bubbles~\cite{Su:2010qj} are included.
\footnote{The morphology of the Bubbles is uncertain, with several alternative templates existing in the literature (e.g.~\cite{Fermi-LAT:2014sfa, Fermi-LAT:2017opo,Macias:2019omb}). It is straightforward in our pipeline to replace the Bubbles template or augment the Bubbles modeling with additional templates, but for simplicity we use the original template from Ref.~\cite{Su:2010qj} in this work to demonstrate the pipeline.}
Three different diffuse models -- Models `A' and `F' from \citet{Calore:2014xka} and Model `O' from \citet{Macias:2019omb} -- are used to model the Galactic diffuse foreground emission, with the gas-correlated ($\pi_0$+bremsstrahlung, or pib for short) emission and inverse Compton scattering (ICS) emission separately modeled as linear combinations of the corresponding components of the three templates.\footnote{It is also entirely possible to extend the diffuse emission modeling to include a larger range of templates, as done in Ref.~\cite{Ramirez:2024oiw} using a purely Poissonian (i.e.~no point sources) version of the pipeline presented in this work; we use this simplified version for demonstration.} Following~\cite{Buschmann:2020adf}, we allow for a modulation of the large-scale spatial structure of the gas-correlated templates modifying it as
$T^p_{\mathrm{diff}}\rightarrow \left[1 + \sum_{\ell m} {A}_{\ell m} \cdot \operatorname{Re}\left(Y^p_{\ell m}\right) \right]T^p_{\mathrm{diff}}$, where the prior on spherical harmonic coefficients ${A}_{\ell m}$ enforces small deviations from the base template; $p\left({A}_{\ell m}\right) = \mathcal U(-0.05, 0.05)$.
Our fiducial choice includes terms up to $\ell=2$, with $m=0,\dots,\ell$ for non-degenerate real parts.
Five spatial templates for diffuse and PS bulge components are included. These are the `Boxy Bulge', `Boxy Bulge Plus', `X-Shaped Bulge' studied in \cite{mcdermott2023morphology} (labeled BB, BBP, and X respectively), the template from Macias \emph{et al} 2019 \cite{Macias:2019omb} labeled Mac19, and the template from Coleman \emph{et al} 2019 \cite{coleman2020maximum} labeled Col19.

In total, our fiducial model is characterized by 41 parameters of interest, which we now enumerate.
For the \textbf{smooth/Poissonian} emission, we fit
\begin{itemize}
    \item Overall normalizations $S_\text{pib}$ and $S_\text{ICS}$ for the gas-correlated (pib) and inverse Compton diffuse components, with mixing fractions $\vec\alpha_\text{pib}$ and $\vec\alpha_\text{ICS}$ specifying the relative contributions of Models `O', `A', and `F' (6 d.o.f.);
    \item Spherical harmonic coefficients $A_{\ell m}$ (up to $\ell = 2$) that modulate the large-scale structure of the gas-correlated templates (5 d.o.f.);
    \item Normalizations $S_\text{bub}$, $S_\text{iso}$, and $S_\text{psc}$ for the \Fermi bubbles, isotropic, and 3FGL catalog emissions (3 d.o.f.);
    \item Smooth GCE normalization $S_\text{GCE}$ with bulge fraction $f_\text{bulge}^\text{pois}$ and mixing fractions $\vec\alpha_\text{bulge}^\text{pois}$ over the five bulge templates, plus the NFW inner slope $\gamma^\text{pois}$ (7 d.o.f.),
\end{itemize}
totaling 21 parameters. For the \textbf{PS/non-Poissonian} emission, we include two PS populations: Galactic disk PS and GCE PS. We fit
\begin{itemize}
    \item Normalizations $S_\text{ps}^\text{GCE}$ and $S_\text{ps}^\text{dsk}$ (2 d.o.f.);
    \item Scale height $z_s$ and radial scale $C$ for the disk spatial template (2 d.o.f.);
    \item GCE PS bulge fraction $f_\text{bulge}^\text{ps}$ with Dirichlet mixing fractions $\vec\alpha_\text{bulge}^\text{ps}$ and NFW inner slope $\gamma^\text{ps}$ (6 d.o.f.);
    \item Doubly-broken power law SCF with slopes $n_1$, $n_2$, $n_3$ and break parameters $S_{b,1}$ and $\lambda \equiv S_{b,2}/S_{b,1}$ (10 d.o.f.), so that
    \begin{equation}
    \frac{\mathrm{d}N}{\mathrm{d}S} \propto
\begin{cases}
\left(\dfrac{S}{S_{b,1}}\right)^{-n_1}, & S \geq S_{b,1} \\[10pt]
\left(\dfrac{S}{S_{b,1}}\right)^{-n_2}, & S_{b,1} > S \geq S_{b,2} \\[10pt]
\left(\dfrac{S_{b,2}}{S_{b,1}}\right)^{-n_2} \left(\dfrac{S}{S_{b,2}}\right)^{-n_3}, & S_{b,2} > S
\end{cases}
    \end{equation}.
    As discussed previously, the $S_{b,i}$ parameters describe the breaks evaluated at the mean exposure; within the code, they are rescaled by the relevant exposure for each exposure region.
\end{itemize}
This totals 20 parameters for the point source model. The priors for all  41 parameters are summarized in Tab.~\ref{tab:prior}.

\begin{figure}[!ht]
\centering
\includegraphics[width=\linewidth]{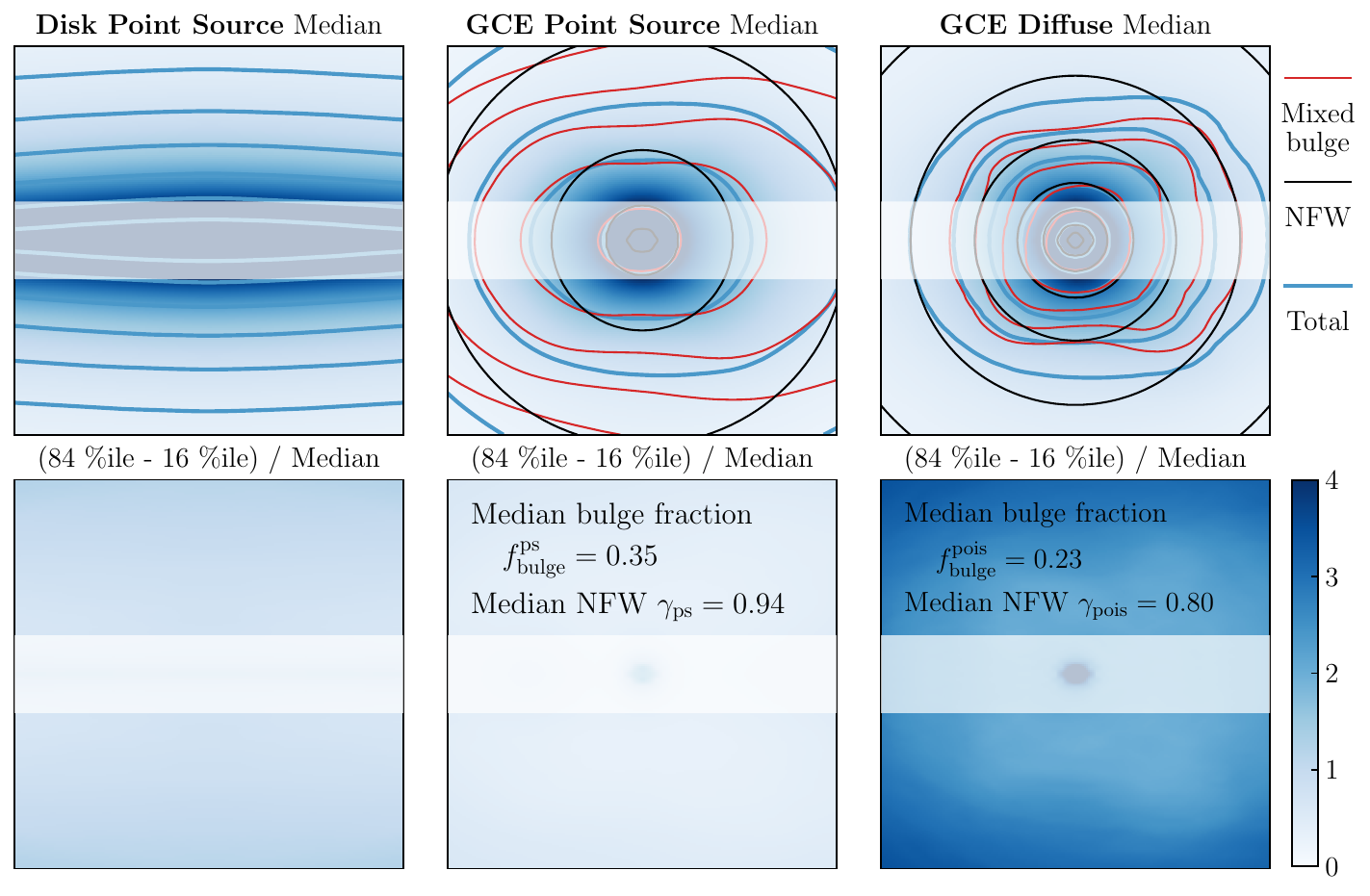}
\caption{Summaries of the inferred posteriors in the inner part of our ROI ($|b|,|l| < 10^\circ$) for disk-correlated PS, GCE-correlated PS, and GCE-correlated smooth emission morphologies from a preliminary analysis of \Fermi-LAT data. The top row shows the median inferred pixel-wise morphologies, while the bottom row shows the relative middle-68\% deviation of the posterior from the median. For comparison, contours for the best-fitting bulge and DM (NFW) description are shown for the GCE PS and smooth components.}
\label{fig:gce-shape}
\end{figure}

\TabPrior

The ROI is taken to be the region within $25^\circ$ of the Galactic Center, masking the Galactic plane at $|b| < 2^\circ$. The template normalization (1 photon per pixel on average) is performed within this full ROI; it is not adjusted for the PS mask. Resolved PSs from the 3FGL catalog are masked at a radius of $0.8^\circ$. \texttt{HEALPix}~\cite{Gorski:2004by} pixelization with resolution parameter $\texttt{nside}=128$ is used, corresponding to pixels of side $\sim0.5^\circ$ and a total of 6839 unmasked pixels in the ROI. 

\subsection{Computation footprint}

For our fiducial 41-parameter model on the full \Fermi dataset, HMC inference required approximately 110 minutes to generate 30,000 samples across 4 chains with 1000 warm-up steps per chain on a single Nvidia A100 GPU. SVI, using $N = 16$ particles to estimate the ELBO gradient at each optimization step, converged in approximately 20 minutes over 5,000 steps on the same hardware, after which 50,000 posterior samples could be drawn in a short time. While both methods produce comparable posteriors (subject to some caveats on the posterior tails in SVI, which we will discuss in Sec.~\ref{sec:validation}), SVI is notably faster for producing a sample size sufficient for characterization of the high dimensional posterior and is expected scale better to higher dimensions \cite{beskos2013optimal, 2012arXiv1206.7051H}.

\subsection{Results}

\begin{figure*}[!ht]
    \centering
    \includegraphics[clip, width=\textwidth]{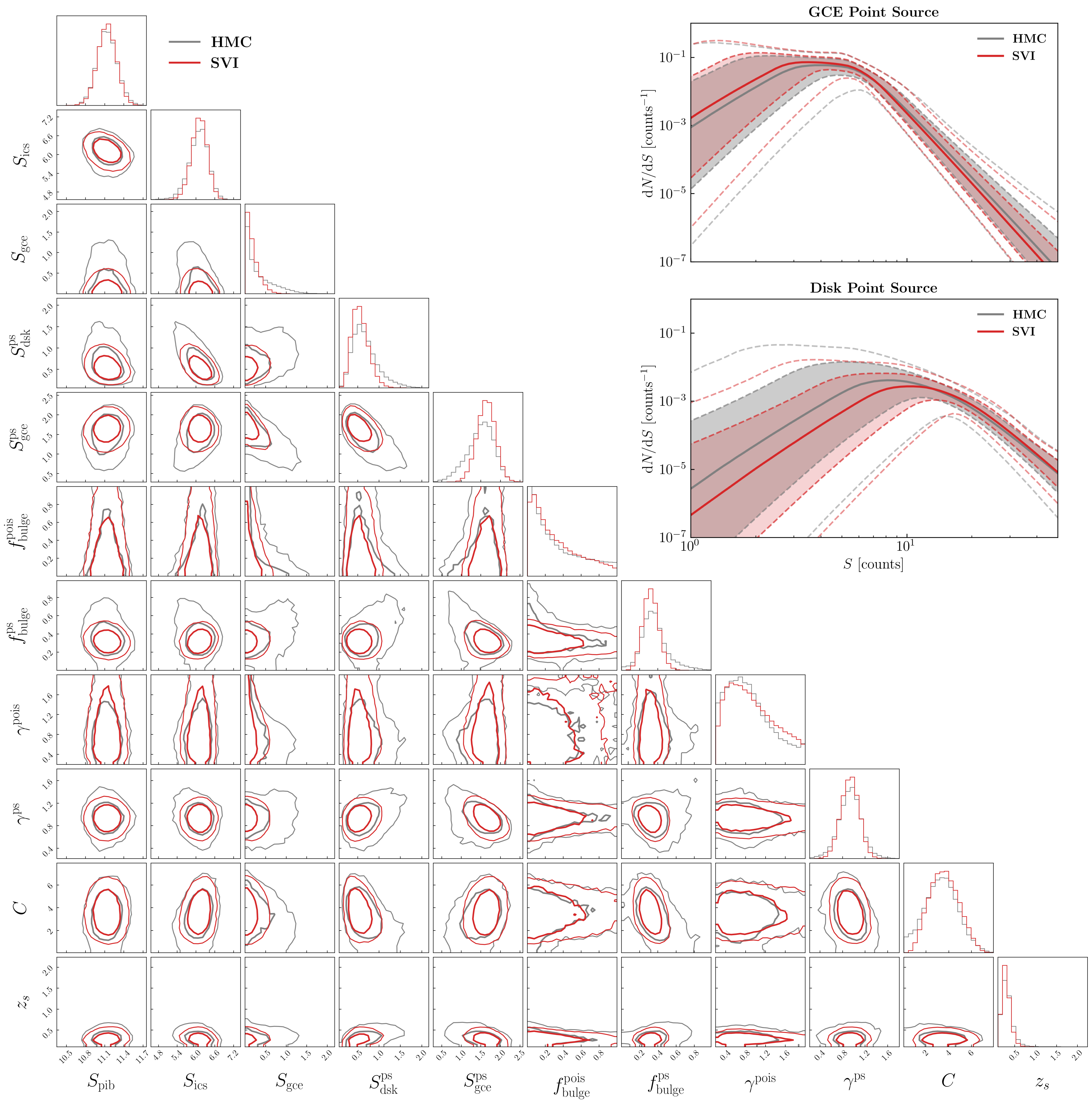}
    \vspace{-0.4cm}
    \caption{{Posteriors of fiducial fit to \textit{Fermi}-LAT data in the Inner Galaxy.} Parameters include normalizations of diffuse components $\pi_0$+bremsstrahlung (pib), ICS, and GCE, where the last is a combination of a mixed bulge template with flux fraction $f_\text{bulge}^\text{pois}$, and a NFW template with flux fraction $1-f_\text{bulge}^\text{pois}$. Also shown are the PS template normalizations for the disk and GCE PS population, where the latter is likewise modeled by a mixture of bulge templates (with overall flux fraction $f_\text{bulge}^\text{ps}$) and a NFW template. The parameters $\gamma^\text{pois}$, $\gamma^\text{ps}$ are the inner slopes for the NFW templates, and $C$ and $z_s$ are scale radius and scale height of the disk PS template. In the corner plot, the inner and outer contours corresponds to 1 and 2 sigma errors. The two panels on the right show posteriors on the SCFs $\mathrm{d}N/\mathrm{d}S$ for the GCE and disk PS populations. The dashed lines denote the 1 and 2 sigma errors (i.e.~$68\%$ and $95\%$ containment) at each $S$ value independently. In all panels, red corresponds to SVI posteriors, and gray corresponds to HMC.
    }
    \label{fig:fermi-post}
\end{figure*}

Figures~\ref{fig:gce-shape} and \ref{fig:fermi-post} summarize the initial results obtained using our pipeline applied to Inner Galaxy \Fermi data. We caution that these results may still suffer from systematics associated with mismodeling of the background and/or signal; we will discuss caveats associated with mismodeling and overconfidence in Sec.~\ref{sec:validation}, and refer the reader to \cite{Ramirez:2024oiw} (which employs a Poissonian version of this pipeline) for a more complete analysis of systematics in the GCE morphology associated with variations in the background modeling. However, these results do demonstrate that the pipeline runs quickly on real data with a high-dimensional space of models, and produces converged results that appear physically reasonable. We will employ these results to define a  realistic ``ground truth'' when validating our pipeline on simulated data.

Summaries of the inferred posteriors for disk-correlated PS, GCE-correlated PS, and GCE-correlated smooth emission morphologies are shown in Fig.~\ref{fig:gce-shape}. The top row shows the median inferred pixel-wise morphologies, while the bottom row shows the relative middle-68\% deviation of the posterior from the median (heuristically, the `1-$\sigma$' range of the posterior distribution). Contours for the best-fitting bulge-correlated (red lines) and DM-like/NFW (black lines) components are shown.

The preferred bulge morphology is noticeably different for the PS and diffuse emission (see Fig.~\ref{fig:fermibulge-post} in the appendix for the triangle plot); in the PS case, there is a preference for the Col19 bulge, whereas in the diffuse case, the bulge-like component is sufficiently small that the fractions of the various components are largely determined by the prior (which is uniform across the various components). This leads to the more boxy ``diffuse bulge'' morphology shown in the right panel of Fig.~\ref{fig:gce-shape}, which is also closer to the NFW profile near the Galactic Center.

Figure~\ref{fig:fermi-post} shows posteriors on select parameters characterizing the GCE and disk PS morphologies. The appendix supplements these results with additional triangle plots. The bulge fractions $f_\mathrm{bulge}^\text{ps}\sim 0.3$ and $f_\mathrm{bulge}^\text{poiss}\sim 0.2$ quantitatively substantiate a modest preference for a NFW-like morphology in the GCE emission. Posteriors on the NFW inner slopes $\gamma$ for the PS and smooth NFW profiles, as well as select parameters characterizing the disk morphology are shown. A physically-reasonable scale height for the disk-correlated PS population, $z_s\sim 0.4$\,kpc is inferred~\cite{bartels2018bayesian}. 

Overall, the posterior median fraction of the GCE emission (including both bulge and NFW components) assigned to unresolved PSs is $88\%$, but the $95\%$ containment region extends up to $100\%$ and down to $41\%$. The inferred SCF for the GCE point sources peaks around 3-5 photon counts per source.\footnote{We have checked that this is not an artificial consequence of the prior on $S_{b,1}^\text{GCE}$, but persists if we extend this prior to lower count values.} These results are similar to those of Ref.~\cite{List:2021aer}, which used Model O for the Galactic diffuse emission and employed a similar ROI and energy window. We caution that systematics in the signal/background modeling may be artificially inflating this fraction \cite{Leane:2019xiy,Leane:2020nmi,Leane:2020pfc}, and also that faint PSs are formally degenerate with diffuse emission; this result should not be seen as a rebuttal to other papers that infer a lower PS fraction (e.g.~\cite{List:2025qbx}), as the main focus of our work is presenting a new pipeline rather than exploring these systematic effects. We note that the diffuse galactic emission model prefers to be predominantly attributed to Model O (Fig.~\ref{fig:fermipibics-post}), consistent with other studies that have found this model provides a better fit to the data compared to the older models A and F (e.g.~\cite{Macias:2019omb,Buschmann:2020adf}).

For both PS and diffuse emission (with this background model and choice of ROI), the inferred posterior prefers a GCE where the majority of the flux is NFW-like, and the bulge component is subdominant. However, it can be seen from Fig.~\ref{fig:gce-shape} that even a small bulge fraction in terms of overall flux can significantly influence the apparent shape of the GCE in the inner Galaxy, especially in the PS case where both the bulge ratio is modestly higher and the Col19 bulge dominates the bulge fraction. The NFW profile extends to higher latitudes than the preferred bulge template, and there is significant flux attributed to the GCE in this region, but within 10 degrees of the Galactic Center, the fluxes associated with the bulge and NFW components are often comparable. We emphasize that we obtain a full posterior distribution over preferred signal morphologies, departing significantly from the rigid parametric descriptions used in older works (although more flexible morphologies have already been considered using an adaptation of this pipeline to the Poisson-only case \cite{Ramirez:2024oiw}).

\section{Validation using simulated data}
\label{sec:validation}

\subsection{Simulated data}

To validate our pipeline, we simulate a large number of skymaps where the ground truth for the parameters is either drawn from the prior, or held at fixed values. The simulated maps are generated using the forward model described in Sec.~\ref{sec:methodology}. 

As mentioned above, the NPTF likelihood implemented in our pipeline is approximate in the case where the telescope has a non-trivial point spread function (PSF). The PSF spreads out photons from a given point source, thus inducing correlations between photon counts in different pixels and violating the approximation that the likelihood can be factorized into individual pixel contributions. It has already been noted in the literature that this approximation can induce overconfidence \cite{Collin:2021ufc}. To study the importance of this effect, we will perform simulations with both a realistic PSF and an artificially narrow PSF whose width is effectively zero (hereafter denoted ``trivial PSF''). When performing parameter inference on simulated data, we include the appropriate PSF corrections (for the simulated map in question) in the NPTF likelihood as discussed in Refs.~\cite{Lee:2015fea, Mishra-Sharma:2016gis}. In our analyses of real data discussed in the previous section, we employed the realistic PSF in the parameter inference.

In the realistic PSF case, we use the King-function model recommended by the \Fermi-LAT collaboration \cite{Fermi-LAT:2013mql}, using the parameters provided in the \texttt{P8R3} instrument response functions for our event selection (\texttt{ULTRACLEANVETO} + top quartile by PSF), at an energy of 2 GeV. These parameters may also be found in the \texttt{NPTFit} code package \cite{Mishra-Sharma:2016gis}. Note that when simulating point sources, we do not employ the (approximate) NPTF likelihood but instead directly sample the source position from the relevant template, and then draw the associated photon positions from the appropriate PSF.

%During the validation we observed that some properties are not stable with respect to variation of the ground truth. Accordingly, we will show results for simulations with two different choices of the ground truth, described in Tab.~\ref{tab:fixed-truths}.

\subsection{Review of relevant statistical coverage tests}

One important set of tests we will perform deals with the statistical coverage properties of our pipeline. SVI in particular is known to be prone to overconfidence, in the context of posterior inference \cite{hermans2022crisis}. In this section we review the tests we will use to identify such overconfidence.

Suppose we seek to infer the value of a scalar quantity $x$. Given a measurement $x_i$ and a desired confidence level $p$, we can construct confidence intervals for $x$, such that (over many experiments) the probability of the truth value being inside the interval is $p$. However, if the likelihood used to construct these confidence intervals is only an approximation to the truth, then the long-run probability of the truth value being inside the interval may instead be $p' \neq p$, in which case we say the \emph{actual coverage} ($p'$) does not match the \emph{nominal coverage} $p$. The curve obtained by plotting $p$ against $p’$ is the coverage curve. To construct this curve, we examine a number (in practice 100) of simulated realizations for a fixed ground truth. For each realization, we apply our fitting pipeline and calculate the confidence intervals for each parameter, for a range of choices of $p$; we then compute the fraction of simulations in which the underlying truth fell within this interval to determine $p'$.

We can also construct a similar coverage curve using Bayesian credible intervals instead of frequentist confidence intervals. In this work we will use highest posterior density intervals on 1D marginal distributions. The nominal coverage is now replaced with the posterior probability coverage $p$. In repeated experiments, the fraction of these credible intervals containing the truth is the actual coverage $p’$. We employ a Bayesian rather than frequentist approach as this is the framework employed by most of the NPTF literature.

Note that while a deviation from $p=p'$ can arise from overconfidence, i.e. a mis-estimation of the shape of the posterior as one moves away from the peak, a {\it bias} in the reconstructed parameters will generally also manifest itself in this way (i.e.~the bias means that the reconstructed credible interval will be systematically shifted away from the truth value, so $p'$ generally lies below $p$). The coverage curve thus provides a diagnostic for both bias and mis-estimation of uncertainties.

For any finite number of simulations, of course, we do not expect perfect agreement between $p$ and $p'$, even with well-calibrated inference. We calibrate the expected scatter in the relation by considering an idealized Gaussian likelihood (with $\sigma=1$ and centered on zero) for a single scalar quantity $x$. The shape of the likelihood does not matter, as long as the confidence intervals are consistently chosen. Specifically, we draw a number of values for $x$ (with probability weighted by the likelihood) equal to the number of simulations we use to construct the real coverage curve (100 in our case), and for each $x$, we compute the minimum value of $p$ for which zero is within the confidence interval for $x$ with confidence level $p$ (given by $p=\text{erf}(|x|/\sqrt{2})$). This provides a list of 100 $p$ values, which we can sort in ascending order to construct a coverage curve (with the corresponding $p'$ values being $1/100, 2/100, \dots, 99/100, 100/100$). We repeat this procedure a large number of times (typically 10,000), and then for each value of $p'$, we determine the $2.5\%$ and $97.5\%$ quantiles for $p$.

We will quantify coverage in two complementary kinds of coverage tests: In the \emph{marginal coverage} tests, we sample the ground truth of each simulation from the prior of the model, so that the actual coverage is marginalized over the prior. In \emph{fixed-truth} (or \emph{conditional}) \emph{coverage} tests, we compute the credible interval coverage evaluated at a single, fixed ground truth. We note that Bayesian credible intervals are guaranteed to be well-calibrated, i.e. approaching $p'=p$ in the limit of exact inference and many realizations, only in the marginal tests \cite{cook2006validation, talts2018validating} where $\theta\sim p(\theta)$, $x\sim p(x\mid\theta)$. At fixed ground truth, no such guarantee exists and the coverage curve may depart from the diagonal even with the exact likelihood and a well-behaved sampler. Nevertheless, it has been pointed out \cite{alokda2026coverage} that marginal calibration may not be entirely sufficient, as a posterior miscalibrated in a truth-dependent way can still cancel when averaged over the prior, and that the fixed-truth coverage may further reveal mis-calibration. We show two sets of \emph{fixed-truth} tests where the ground truths are selected to be reasonably close to the \Fermi data fit, and show that the coverage behavior can indeed vary with the choice of ground truth values. We first present the marginal results in Sec.~\ref{subsec:marginal-coverage}, then the fixed-truth results in Sec.~\ref{subsec:fixed-truth-coverage}.

\subsection{Marginal coverage}
\label{subsec:marginal-coverage}

\begin{figure}[ht!]
    \centering
    \includegraphics[width=\linewidth]{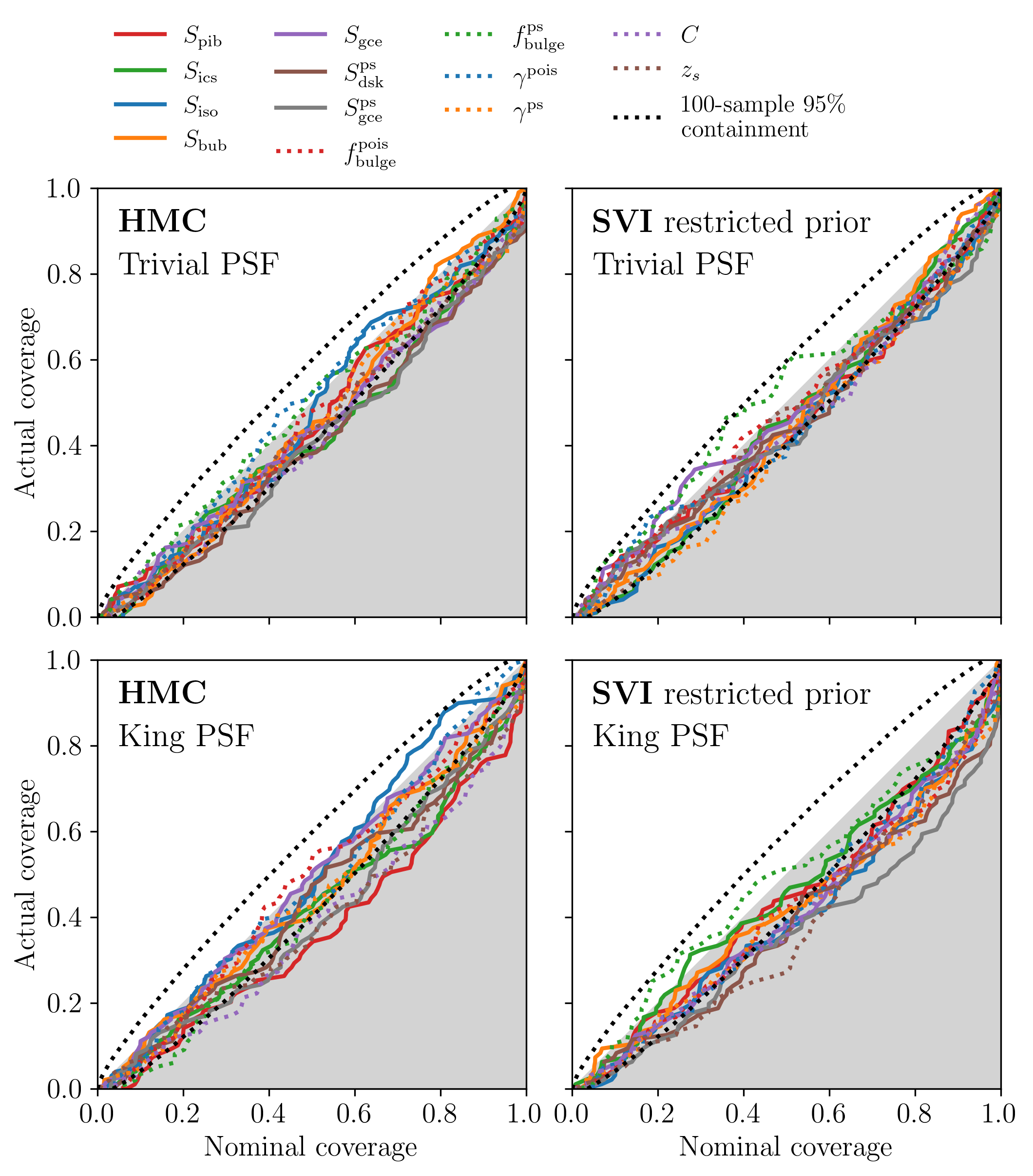}
    \caption{Marginal coverage curves generated with 100 simulations, whose ground truths are drawn independently from the prior. Colored lines correspond to different parameters; dotted lines correspond to the expected $95\%$ containment region for a Gaussian likelihood. Only a representative set of parameters are shown.}
    \label{fig:prior-coverage}
\end{figure}

We begin with the marginal coverage test, in which the ground truth of each of the 100 simulations is drawn independently from the prior, shown in Tab.~\ref{tab:prior}. The coverage curves obtained from HMC (left column) and SVI (right column) posteriors are shown in Fig.~\ref{fig:prior-coverage}. The top row corresponds to the case where both the simulated data and the NPTF likelihood assumes a trivial PSF, while the bottom row employs the realistic King PSF.

For HMC, the coverage curves lie within the expected scatter band for the trivial PSF case (for which the NPTF likelihood is exact). The good marginal coverage confirms that the NUTS HMC sampler and our pipeline implementation are unbiased and correctly calibrated in the marginal sense. In the King PSF case, the coverage curves are slightly biased as expected, since the NPTF likelihood is no longer exact in the presence of unresolved point sources.\footnote{We plan to explore and further characterize the fact that marginal posteriors of some normalization variables can be overconfident under the approximate NPTF likelihood in a future work.} Despite this, the King PSF coverage remains approximately well-calibrated.

For SVI, a practical complication forces us to use a prior with reduced range: the memory footprint of recursive NPTF likelihood evaluation scales with the maximum photon count appearing in any pixel. Ground truths drawn from the full prior occasionally have large normalizations, producing high photon counts ($\gtrsim300$) that exhaust GPU memory under the fiducial SVI setup and prevent some of the 100 fits from completing. Given a specific hardware configuration and the default \texttt{numpyro} implementation of the SVI inference loop, one can always reduce the IAF size to accommodate the high photon numbers. However, since the real \Fermi data can be accommodated, and we believe the choice of prior range matters little to the coverage test (as we will show), we decide to maintain the fiducial architecture for SVI. We choose to run the SVI marginal coverage test using a \emph{restricted} prior, with more stringent upper bounds on the component normalizations that reduce the maximum photon counts while remaining wide enough to encompass the \Fermi\ data. Other than this complication, the SVI marginal coverage tells a similar story as the HMC, with the trivial PSF case being well-calibrated, and the King PSF case showing a slight overconfidence, due to the likelihood being approximate.

We stress that the restricted prior is only used in the SVI marginal coverage test, and our fiducial analysis retains the full prior range of Tab.~\ref{tab:prior}. The memory bottleneck we encountered is set by the maximum photon count in the data, not by the prior, and the \Fermi data we use is comfortably within reach of our SVI setup. To verify that the restricted prior is viable, we fit the \Fermi data with the restricted prior and recover essentially identical posteriors (see Fig.~\ref{fig:fermi-post-restricted} in App.~\ref{appd:restricted-prior}). We leave improvements to the memory scaling of the high-photon-count regime to future work. As a cross-check that the restricted prior is a fair surrogate, we also ran the marginal coverage test for HMC with the restricted prior; the results (Fig.~\ref{fig:marginal-coverage-hmc-restricted} in App.~\ref{appd:restricted-prior}) remain well-calibrated and consistent with the full prior-range HMC results.

While the marginal coverage is promising, we caution the reader that truth-dependent miscalibration can be present. A warning sign is visible in the fit to real data in Fig.~\ref{fig:fermi-post}: SVI tends to produce more confident posteriors for certain variables, indicating a potential issue with at least one of the sampling methods. For a more localized picture, we turn to a fixed-truth coverage probing the specific region most relevant to the \Fermi data.

\subsection{Fixed-truth coverage}
\label{subsec:fixed-truth-coverage}

We now describe fixed-truth coverage tests performed on the same pipeline, with two ground truths $A$ and $B$, chosen so that (1) they are close to the \Fermi posterior, or simple point estimates to the \Fermi data; and (2) they exhibit different coverage performance, where fits on simulations generated with truth $A$ show better coverage properties than those with truth $B$. The truths are summarized in Tab.~\ref{tab:fixed-truths}.

\begin{figure*}[ht!]
    \centering
    \includegraphics[width=\linewidth]{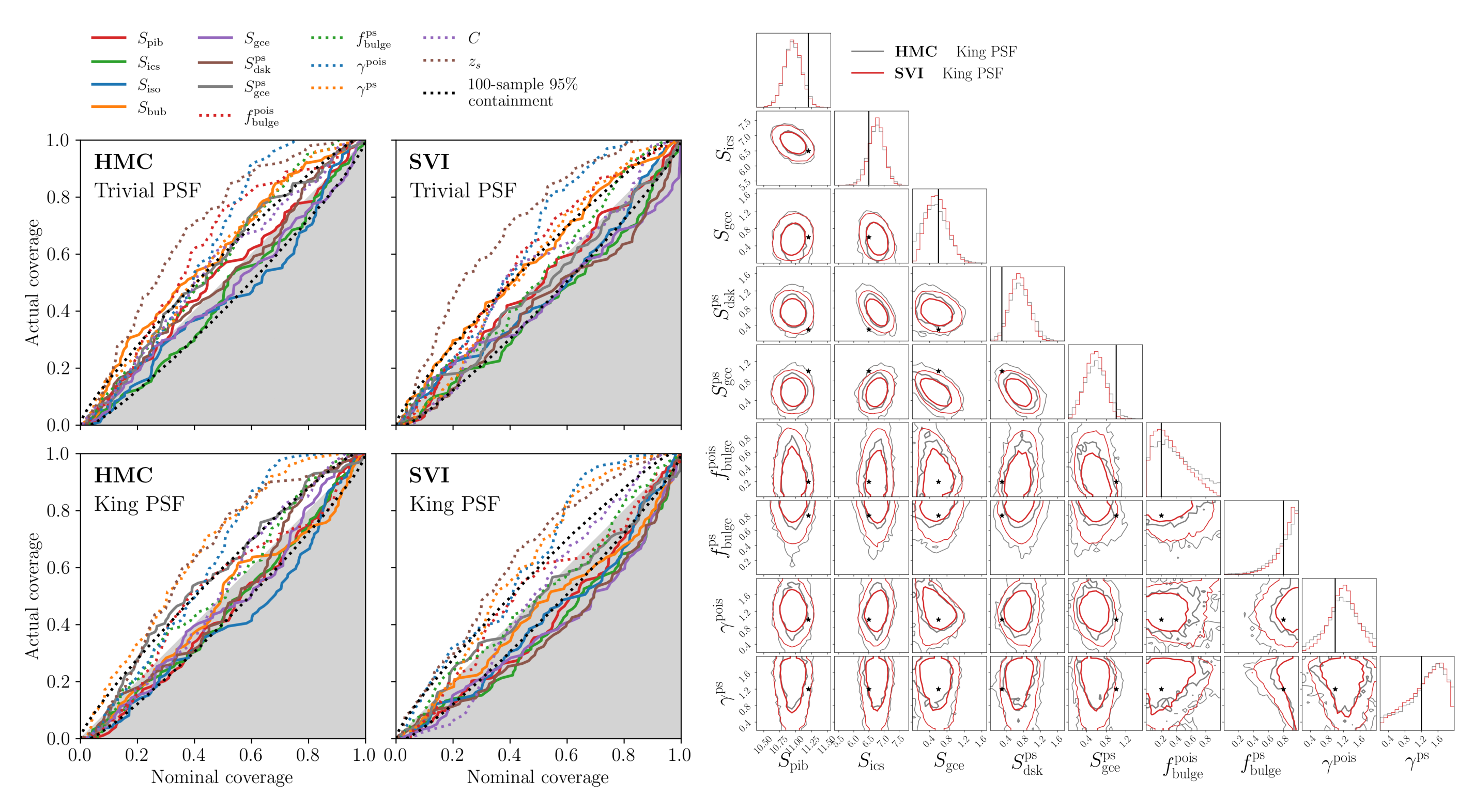}
    \caption{Coverage curves (left) and example posteriors (right) for ground truth $A$ (see Tab.~\ref{tab:fixed-truths}). Coverage curves are generated using 100 simulations; colored lines correspond to different parameters, dotted lines correspond to the expected $95\%$ containment region for a coverage curve with 100 samples. For the posteriors, a realistic King PSF is employed. Only a representative set of parameters are shown.}
    \label{fig:fixed-truth-example-A}
\end{figure*}

We first show in Fig.~\ref{fig:fixed-truth-example-A} the coverage for truth $A$. As before, in the trivial PSF case, the SVI and HMC posteriors appear to match closely and to be quite well-calibrated, with perhaps a tendency for some parameters to be mildly overconfident and others underconfident in the case of SVI with a King PSF. Example posteriors for one of the simulated data realizations are shown in Fig.~\ref{fig:fixed-truth-example-A}, and we observe that the posterior appears unimodal. We have checked that this is a common feature among posteriors in this simulation suite.

\TabTruth

\begin{figure*}[ht!]
    \centering
    \includegraphics[width=\linewidth]{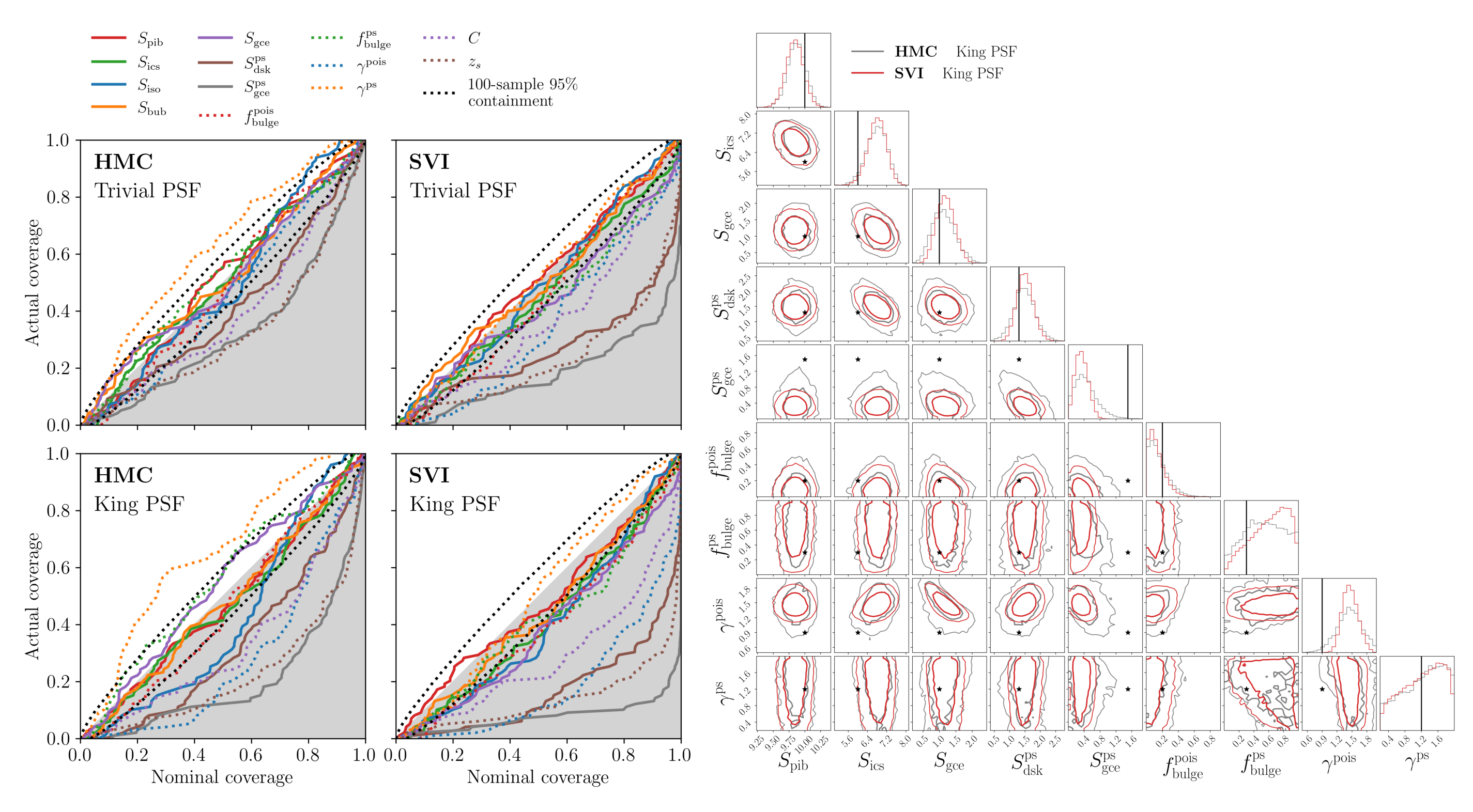}
    \caption{Coverage curves (left) and example posteriors (right) for ground truth $B$ (see Tab.~\ref{tab:fixed-truths}). Coverage curves are generated using 100 simulations; colored lines correspond to different parameters, dotted lines correspond to the expected $95\%$ containment region for a coverage curve with 100 samples. For the posteriors, a realistic King PSF is employed. Only a representative set of parameters are shown.}
    \label{fig:fixed-truth-example-B}
\end{figure*}

However, this good behavior is not ubiquitous, even for relatively modest changes in the ground truth. We show analogous results in Fig.~\ref{fig:fixed-truth-example-B}, which is based on a simulation suite with ground truth $B$. We observe that in this case, some overconfidence is already apparent even in the trivial PSF case for which the NPTF likelihood is exact, and is more pronounced for the SVI analysis than for HMC. The overconfidence then becomes more severe in the realistic King PSF case. This is consistent with the picture of Sec.~\ref{subsec:marginal-coverage}, since deviations from $p'=p$ at a single fixed ground truth are permitted and do not contradict the good marginal coverage. However, the large degree of overconfidence reveals a localized drawback of our pipeline. In the following, we attempt to trace its origin.

\begin{figure}[ht!]
    \centering
    \includegraphics[width=\linewidth]{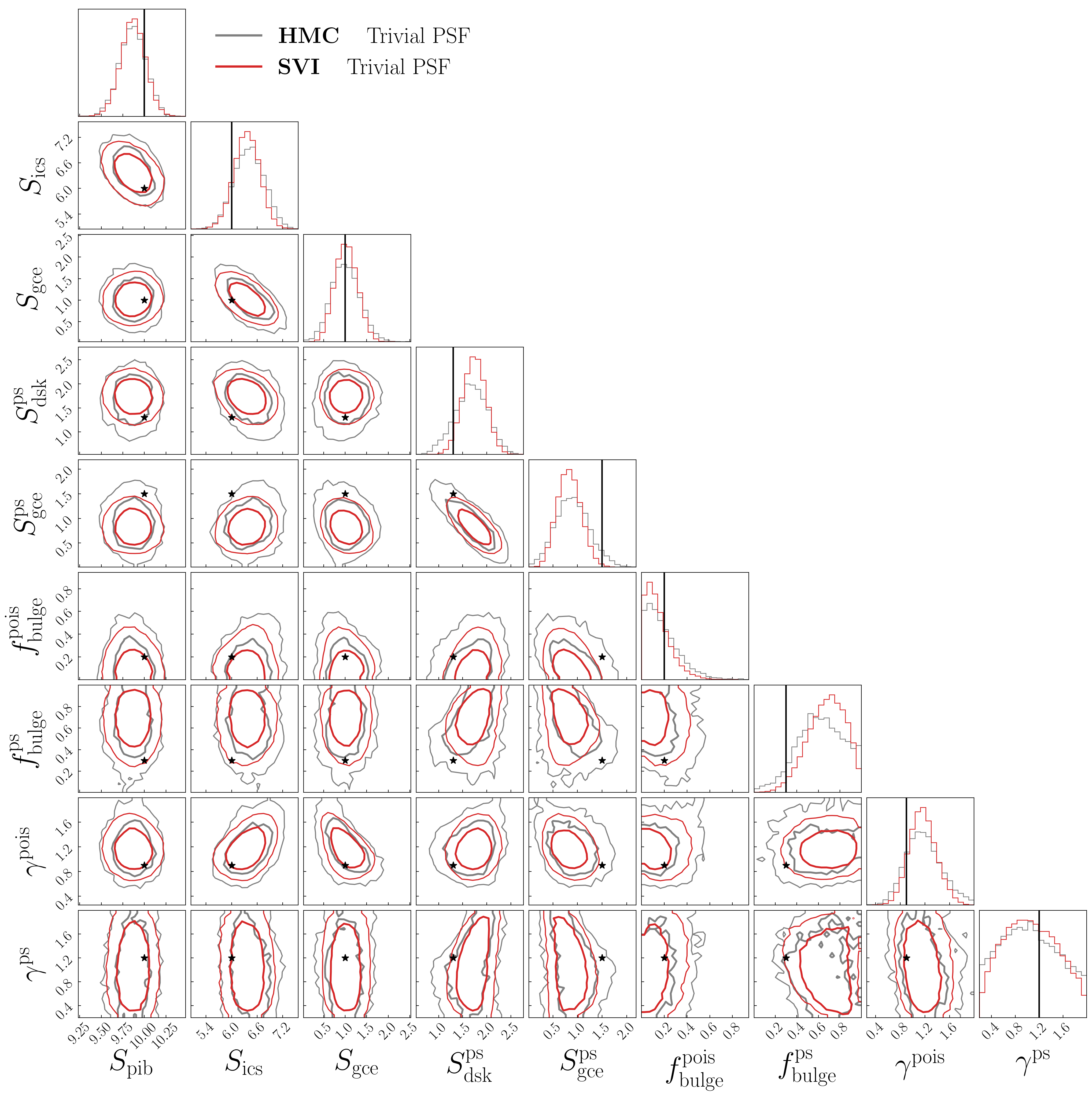}
    \caption{Example posteriors for the ground truth $A$, in the case with a trivial PSF. Only a representative set of parameters are shown.}
    \label{fig:npold-post-deltapsf}
\end{figure}

To understand the source of the problem, we can compare the shape of the HMC posterior between the two examples in Fig.~\ref{fig:fixed-truth-example-A} and Fig.~\ref{fig:fixed-truth-example-B}. We observe that the true value of the parameter often lies outside of the mode characterized by the maximum likelihood in the miscalibrated example. In all the cases we have checked, a comparison of likelihood values always shows that the likelihood of the truth value and that of the HMC best fit (maximum a posteriori) value are very similar. This indicates that the true posterior of the NPTF fit, for this ground truth, is multi-modal (at least for an appreciable fraction of realizations), and the difference between these modes are small. In these circumstances, the approximation in the likelihood may have a greater impact than otherwise, potentially skewing the (approximate) maximum likelihood point to a different region in parameter space (away from the truth value). More generally, even when the likelihood is exact, obtaining the correct multi-modal posterior in a high-dimensional fit is challenging. We observe that the posterior is not multi-modal in the well-calibrated example of Fig.~\ref{fig:fixed-truth-example-A}. In the trivial PSF case with the miscalibrated example (with posteriors shown in Fig.~\ref{fig:npold-post-deltapsf}), there is discrepancy between the SVI and HMC posteriors, but it is less severe compared to the case with a realistic King PSF.

The overconfidence is noticeably worse in the SVI case vs HMC. SVI with the reverse KL divergence is known to be mode-seeking~\cite{li2016renyi}, i.e. tending to concentrate the posterior around a single mode of the true posterior at the expense of underrepresenting other modes and producing miscalibrated posterior tails. While the normalizing flow used as our variational family is in principle expressive enough to represent multi-modal distributions~\cite{rezende2015variational}, the mode-seeking nature of the reverse KL divergence means that in practice the flow can collapse onto a single mode. This is consistent with what we observe in Fig.~\ref{fig:fixed-truth-example-B}: the SVI posterior can fail to reproduce the complicated, potentially multi-modal posterior shapes found by HMC, particularly for parameters where the ground truth lies in a secondary mode. We suspect that the high dimensionality of the parameter space and the shallowness of the likelihood surface exacerbate this mode-seeking behavior, making it harder for the variational optimization to explore the full posterior landscape.

We have attempted a few known fixes to this issue, including (1) fitting with a more expressive IAF; (2) changing the KL-divergence in the SVI loss function to a R\'enyi divergence~\cite{li2016renyi}, with $\alpha$ values that makes the inference optimization more mode-covering; (3) replacing the Gaussian base distribution of the normalizing flow with a heavier-tailed Student's $t$ distribution; (4) tempering the likelihood by annealing a multiplicative factor $\beta$ on the log-likelihood from 0 to 1 during optimization~\cite{mandt2016variational, huang2018improving}; (5) changing the learning rate schedule and optimizer. However, while incremental progress can be observed (\textit{e.g.} a small value of the R\'enyi $\alpha$ does broaden the SVI posterior tails --- but the coverage test remains miscalibrated) none of these changes completely address the issue. For the moment, we recommend validating final SVI-based results with HMC, especially if the SVI posterior appears multimodal.

Even the HMC may suffer from some difficulty in fully mapping out the multi-modal posterior. Where there is evidence of multi-modality (from the HMC posterior directly, or from substantial discrepancies between the HMC and SVI posteriors), methods such as parallel tempering \cite{PhysRevLett.57.2607,geyer1991markov} can be used to try to map out the full multi-modal structure, which we have implemented in the public pipeline.

\begin{figure*}[ht!]
    \centering
    \includegraphics[width=0.8\linewidth]{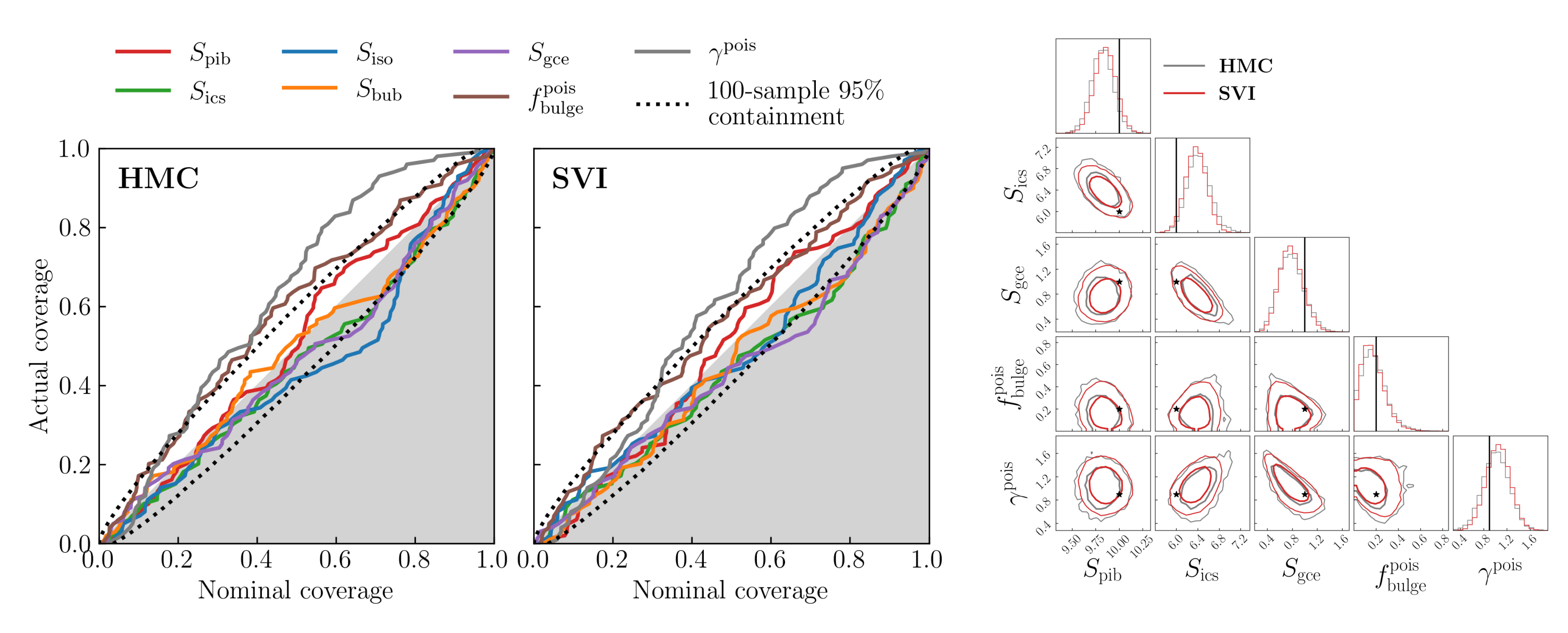}
    \caption{Coverage curves (left) and example posteriors (right) for the ground truth $B$, but restricted to the Poissonian templates. Coverage curves are generated using 100 simulations; colored lines correspond to different parameters, dotted lines correspond to the expected $95\%$ containment region for a coverage curve with 100 samples. Only a representative set of parameters are shown.}
    \label{fig:pois-coverage}
\end{figure*}

We add that posteriors generated by both the HMC and SVI are well-calibrated in a reduced model without point source components. In this case, the dimensionality of the model is reduced, and an exact Poissonian likelihood can be used. While we also confirmed that the marginalized coverage is well-calibrated, we present the coverage curve and example posterior constructed with truth $B$ in Fig.~\ref{fig:pois-coverage}; this truth previously exhibited problematic coverage when point source components were included. The Poissonian version of the pipeline has been extended and modified to study the morphology of the GCE in a template-independent way using a Gaussian process model for the signal \cite{Ramirez:2024oiw}; those results should not be affected by the overconfidence we have identified in the presence of source populations.

\subsection{Cross check with \texttt{NPTFit}}

\begin{figure}[ht!]
    \centering
    \includegraphics[width=\linewidth]{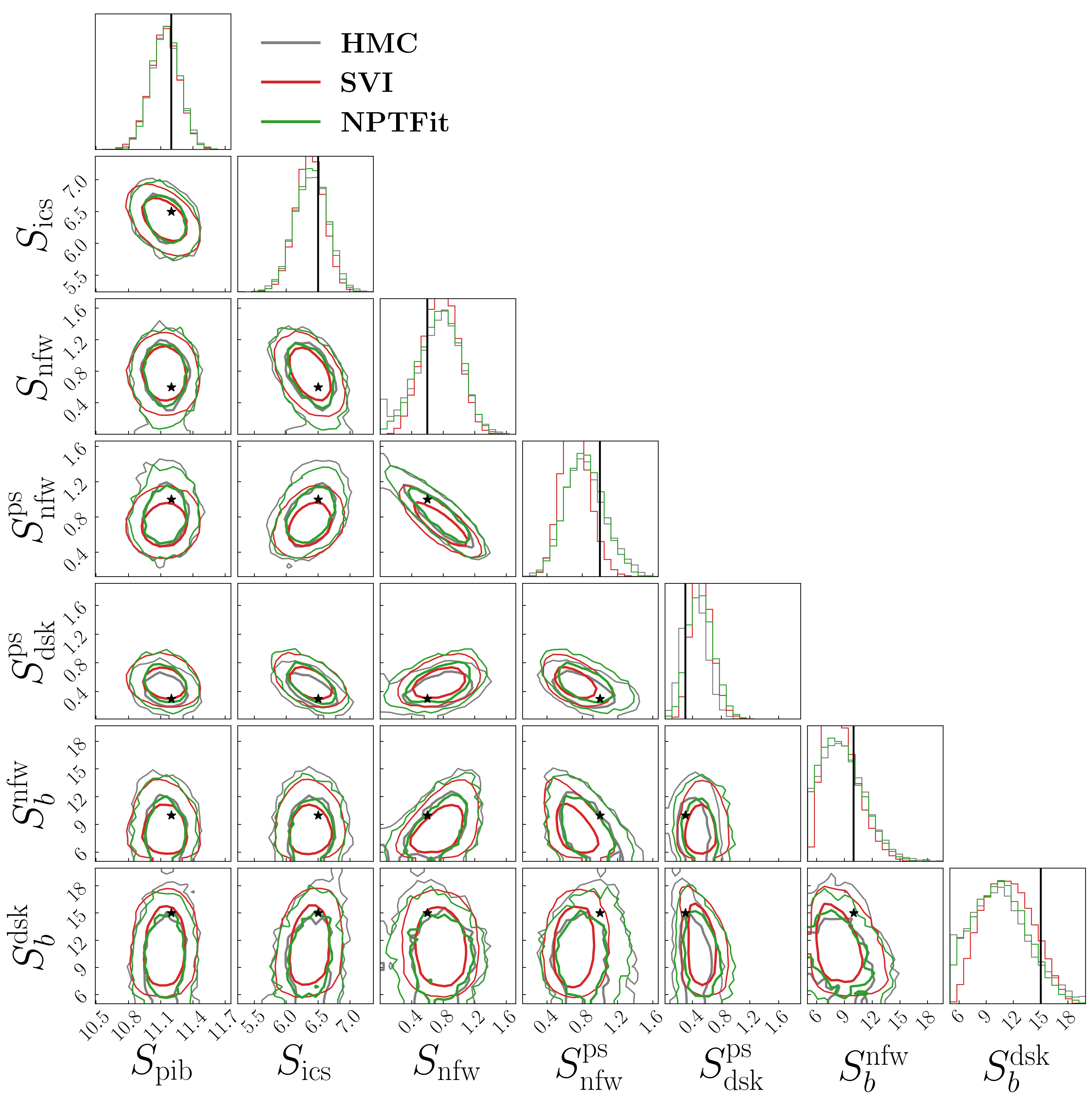}
    \caption{{Posterior comparison between the differentiable pipeline and \texttt{NPTFit} and ground truth using a 14-parameter model.} Only a representative set of parameters are shown.}
    \label{fig:cmp}
\end{figure}

As a cross check, we compare the posteriors obtained using our differentiable pipeline against those from the original \texttt{NPTFit} code package~\cite{Mishra-Sharma:2016gis}, which uses \texttt{MultiNest}~\cite{Feroz:2013hea} as its sampling backend. Since \texttt{NPTFit} is restricted in the number of parameters in a fit, we reduce the model for this comparison to contain smooth templates for the $\pi_0$+bremsstrahlung (pib) and ICS diffuse components, the \Fermi bubbles, isotropic emission, and resolved PSs from the 3FGL catalog, along with a smooth NFW component with the inner slope parameter fixed to $\gamma = 1.0$, totaling 6 templates. Two PS populations are included: a disk component and a NFW component with a steeper inner slope of $\gamma = 1.2$. To ensure a consistent comparison with \texttt{NPTFit}, we match our differentiable model to use single-break power laws for the SCF and uniform priors on the multiplicative coefficient $A$ of the SCF, rather than the count normalization $S_\mathrm{PS}$ used in our fiducial analysis. Together with a break location $S_b$ and two slopes $n_1$ and $n_2$ for each PS component, the comparison model totals 14 parameters. We generate simulated data from this reduced model to compare posteriors, with the ground truth values shown in Fig.~\ref{fig:cmp} (comparable to those in ground truth $A$).

Fig.~\ref{fig:cmp} compares the posteriors for a subset of parameters, demonstrating good agreement between HMC, SVI, and \texttt{NPTFit}. We note that while the bulk of the posteriors are consistent across the three methods, some variation in the tails are still present; this is not surprising given the sensitivity of the tails to different sampling algorithms.

The differentiable pipeline does offer substantial computational speed advantages. \texttt{NPTFit}, based on \texttt{MultiNest}, is not GPU-accelerated; using 1000 live points distributed across 16 CPUs (Intel Xeon Platinum 8480CL), it produced a final sample of 7272 posterior points in approximately 90 minutes. By contrast, our HMC implementation drew 10,000 samples across 4 chains in 45 minutes on a single Nvidia A100 GPU, while SVI converged in under 5 minutes in around 1000 steps on the same hardware (after which 50,000 posterior samples are drawn in a few seconds). These results demonstrate that the differentiable inference pipeline can produce fast and sensible posteriors, although we again caution the caveats regarding the tails of the posterior distribution in multi-modal cases (especially for SVI) discussed previously.

\begin{figure}[h]
    \centering
    \includegraphics[width=\linewidth]{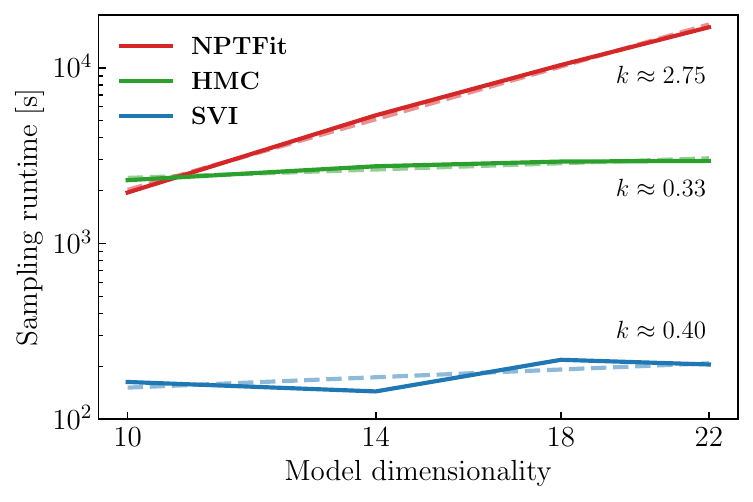}
    \caption{Inference runtime as a function of model dimensionality. Despite limited data, both HMC and SVI exhibit substantially more favorable scaling than \texttt{NPTFit}.}
    \label{fig:scaling}
\end{figure}

We further explore briefly how runtime scales with model dimensionality. We constructed variants of the comparison model from 10 to 22 parameters by removing or adding PS populations: a 10-parameter model obtained by dropping the disk PS component, and 18- and 22-parameter models obtained respectively by adding an isotropic PS component, and another tracing the Col19 bulge, atop the fiducial 14-parameter model. For HMC, we draw 10000 samples across 4 chains, and for SVI, we run optimization until the relative change in the ELBO falls below $10^{-6}$ in a 200-step rolling average; all other settings match those used in the comparison above. Runtime as a function of model dimension is shown in Fig.~\ref{fig:scaling}. \texttt{NPTFit}, using \texttt{MultiNest}, exhibits a scaling exponent of approximately $2.75$, consistent with the $\sim d^{3}$ scaling expected for nested sampling~\cite{chopin2010properties}. NUTS HMC and SVI scale much more gently, with measured exponents of roughly $0.33$ and $0.4$, respectively, below literature expectations of $d^{5/4}$ for HMC~\cite{beskos2013optimal} and $d^{1}$ for SVI~\cite{2012arXiv1206.7051H}. We attribute these differences to the small dimensionality of the models tested, where hardware capacity is not yet saturated. Nevertheless, the qualitative hierarchy illustrates the favorable scaling of gradient-based methods for the high-dimensional parameter spaces studied in this work.

\subsection{Addressing diffuse component mismodeling via a mixture of templates}

Our model's ability to accommodate many parameters allow us to bring together multiple templates for the same component in our model. Specifically, for the Poissonian and point source bulge components we use a linear combination of 5 bulge templates, and for pib and ICS, we use a linear combination of Model `O', `A', and `F' with large scale modulation by spherical harmonics up to $l=2$. It is well known that the commonly-used single templates for these components are not sufficient to model the data down to the level of noise and that background and/or signal mismodeling can have substantial effects on the posteriors of interest (e.g.~\cite{Buschmann:2020adf,Leane:2020nmi,Leane:2020pfc}), so the hope is that adding additional flexibility to these components may mitigate the mismodeling and provide more reliable posteriors. In this subsection, we explicitly test this hypothesis in simulated data, focusing on mismodeling of the diffuse background.

\begin{figure*}[ht]
    \centering
    \includegraphics[width=0.9\linewidth]{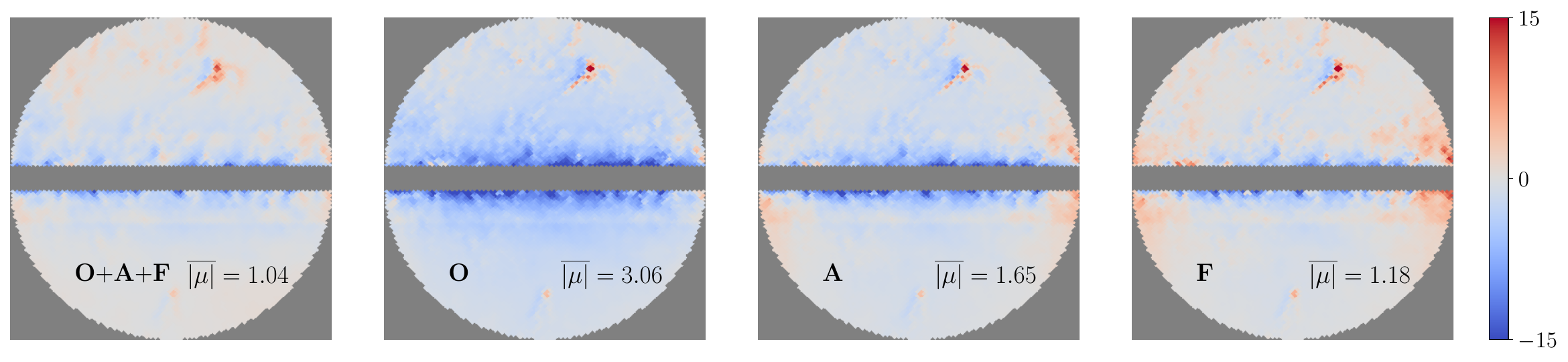}
    \caption{{Differences between median of reconstructed diffuse emission and the truth}, for simulation tests with the \texttt{P6V11} diffuse template, fitted with a linear combination of Model `O' `A' `F', and individual models. We show overlaid on each plot the mean absolute value difference in photon counts per pixel. The truth mean value of \texttt{P6V11} is taken to be 17.7 photons per pixel.}
    \label{fig:diffuse-diff}
\end{figure*}

Our linearly combined templates are set up such that the normalization $S$ is sampled from a uniform distribution, and the fractions of each template $\vec\alpha$ are sampled from a flat Dirichlet distribution. We first test explicitly that in the absence of mismodeling, we can correctly recover the fractions of the different diffuse templates. We simulate, as an example, maps with ground truth $A$ (Tab.~\ref{tab:fixed-truths}), modified to have fractions $\alpha_\text{pib}$ for models `O', `A', and `F' as 0.4, 0.0, and 0.6 respectively. The recovered posterior has $\alpha_{\text{pib, O}}=0.35\pm0.06$, and $\alpha_{\text{pib, F}}=0.58\pm0.08$, while $\alpha_{\text{pib, A}}$ is consistent with 0 with an uncertainty of 0.06.

We then sought to test the effects of mismodeling. We generated simulated data using the \texttt{P6V11} diffuse model released by the {\it Fermi} Collaboration (generated as described in the original NPTFit paper \cite{Lee:2015fea}) to provide the $\pi_0$+bremsstrahlung (pib) and ICS components. This model is known to be markedly different from Models A/F/O at a level sufficient to generate spurious evidence in favor of point sources \cite{Buschmann:2020adf}.  We simulate maps with ground truth $A$ (Tab.~\ref{tab:fixed-truths}) albeit with $S_\text{pib}=S_\text{ics}=0$, and $S_\text{P6V11}=17.7$, the sum of the original $S_\text{pib}$ and $S_\text{ics}$ values (since the \texttt{P6V11} template does not separate the various diffuse components).

\begin{figure}[ht]
    \centering
    \includegraphics[width=\linewidth]{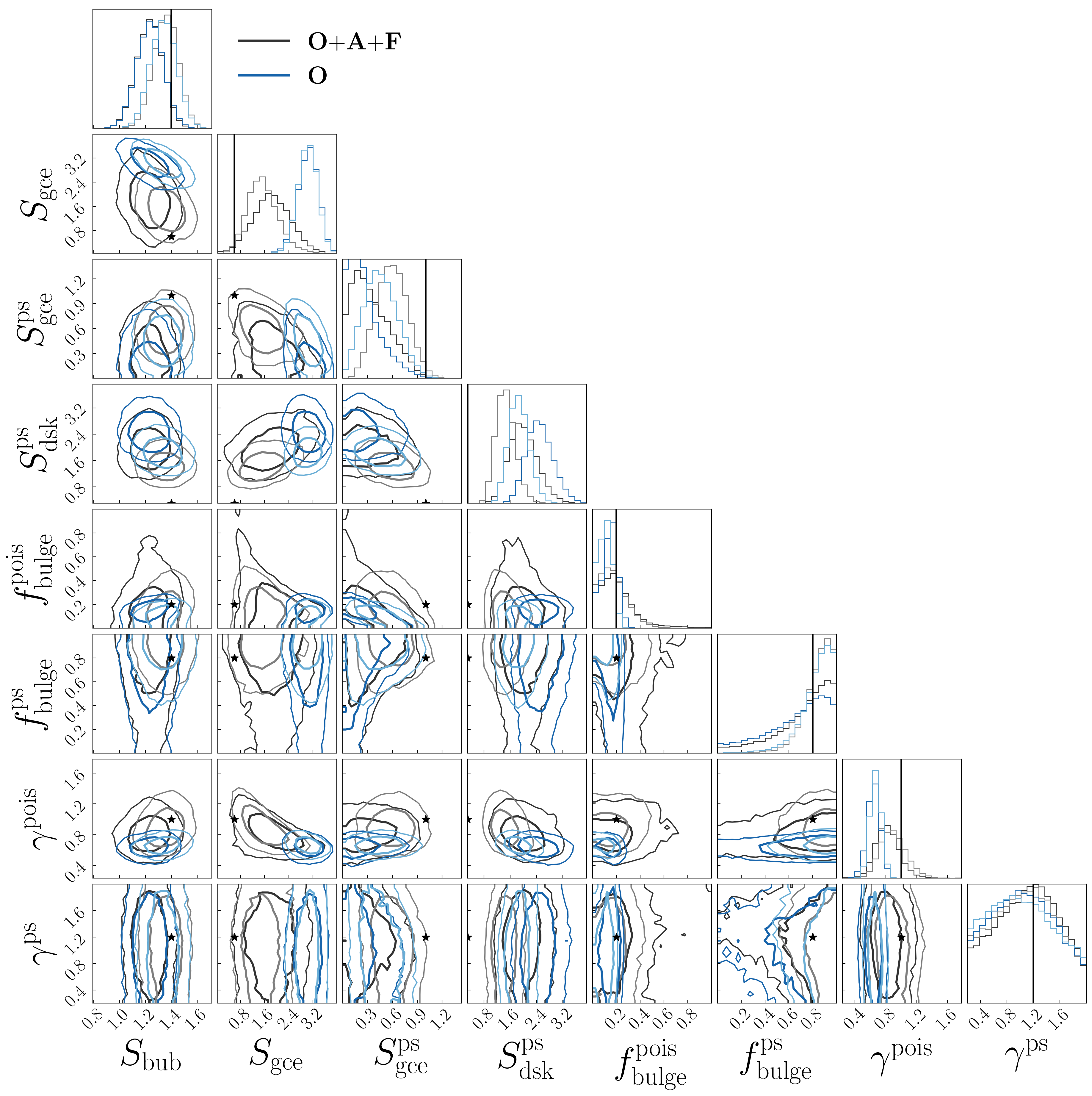}
    \caption{Posterior comparison between fitting with the combined Model `O'+`A'+`F' diffuse template (black and gray) and Model `O' alone (blue), for select parameters using data simulated with the \texttt{P6V11} diffuse template. Posteriors corresponding to two simulated realizations are shown for each case.}
    \label{fig:oafvso}
\end{figure}

We study the ability of the mixture template to mitigate mismodeling, by testing the fit to simulated data with the mixture template vs single templates. Our fiducial model is run, along with three other models with only one of Models `O', `A', and `F' available (with spherical harmonic modulation in place in all cases). The median recovered diffuse component is shown in Fig.~\ref{fig:diffuse-diff}, where the mixture template shows better recovery of the mismodeled diffuse component compared to single template fits, evidenced by a lower mean absolute photon count difference. We also observe that the combined fit yields improved posteriors for non-diffuse parameters compared to fitting with Model `O' alone, in particular for $S_\text{gce}$, as shown in Fig.~\ref{fig:oafvso}. We note, however, that the recovered posteriors are not perfectly calibrated in either case (and there is certainly some degree of residual mismodeling).

\section{Discussion and conclusions} 
\label{sec:conclusion}

We have introduced a scalable and flexible pipeline for disentangling the contribution of various emission components to the observed $\gamma$-ray sky using differentiable probabilistic programming. The pipeline makes extensive use of the seamless vectorization, automatic differentiation, just-in-time compilation, and GPU acceleration enabled by the \texttt{Jax} framework as well as the flexible model specification and inference capabilities of \texttt{NumPyro}. 
As a first example, we applied our pipeline to Inner Galaxy \Fermi-LAT $\gamma$-ray data with the goal of characterizing the longstanding GCE emission while allowing for a richer description of possible signal morphologies and Galactic foreground model descriptions.

There are a number of possible extensions to this pipeline. While we have used parametric and physically-motivated morphologies for the spatial distribution of various PS populations, using our pipeline we will be able to characterize the Galactic Center in a fully signal model-independent manner by scanning over a space of spatial functions on the sky map e.g., using Gaussian processes (GPs) to define the function prior. A similar GP-based technique was implemented in \citet{Mishra-Sharma:2020kjb} to augment the morphology of the diffuse foreground template, and GPs have been used to model the morphology of the GCE signal in an earlier Poisson-only version of our pipeline \cite{Ramirez:2024oiw}. Finally, since our pipeline is implemented using differential programming, it can admit arbitrarily flexible descriptions of the Galactic foreground emission, e.g. as a latent-variable generative model that spans a distribution of possible foreground models. We leave a study of these extensions to future work.

\section*{Acknowledgements}

This work is supported by the National Science Foundation under Cooperative Agreement PHY-2019786 (The NSF AI Institute for Artificial Intelligence and Fundamental Interactions, \url{http://iaifi.org/}).
This material is based upon work supported by the U.S. Department of Energy, Office of Science, Office of High Energy Physics of U.S. Department of Energy under grant Contract Number DE-SC0012567.
This research was supported in part by grant NSF PHY-2309135 to the Kavli Institute for Theoretical Physics (KITP).
During the completion of this work, T.R.S.~was supported in part by a Guggenheim Fellowship; the Edward, Frances, and Shirley B.~Daniels Fellowship of the Harvard Radcliffe Institute; and the Bershadsky Distinguished Fellowship of the Harvard Physics Department.
Y.S.~was supported by a Trottier Space Institute Fellowship.
The computations in this paper were run on the FASRC Cannon cluster supported by the FAS Division of Science Research Computing Group at Harvard University. T.R.S.'s work is supported by a grant from the Simons Foundation (Grant Number 929255, T.R.S). This work was performed in part at the Aspen Center for Physics, which is supported by National Science Foundation grant PHY-2210452.
We thank the \Fermi-LAT Collaboration for making publicly available the $\gamma$-ray data used in this work.
This research has made use of NASA's Astrophysics Data System.
This research made use of the
\texttt{astropy}~\cite{Price-Whelan:2018hus,Robitaille:2013mpa},
\texttt{dynesty}~\cite{Speagle_2020},
% \texttt{getdist}~\cite{Lewis:2019xzd},
\texttt{IPython}~\cite{PER-GRA:2007},
\texttt{Jax}~\cite{jax2018github},
\texttt{Jupyter}~\cite{Kluyver2016JupyterN},
\texttt{matplotlib}~\cite{Hunter:2007},
% \texttt{MLflow}~\cite{10.1145/3399579.3399867},
% \texttt{nflows}~\cite{nflows},
\texttt{NPTFit}~\cite{Mishra-Sharma:2016gis},
\texttt{NPTFit-Sim}~\cite{NPTFit-Sim},
\texttt{NumPy}~\cite{harris2020array},
% \texttt{pandas}~\cite{pandas:2010},
% \texttt{PyGSP}~\cite{michael_defferrard_2017_1003158},
\texttt{NumPyro}~\cite{2019arXiv191211554P},
\texttt{Pyro}~\cite{bingham2019pyro},
\texttt{PyTorch}~\cite{NEURIPS2019_9015},
\texttt{PyTorch Geometric}~\cite{Fey/Lenssen/2019},
% \texttt{PyTorch Lightning}~\cite{william_falcon_2020_3828935},
% \texttt{seaborn}~\cite{seaborn},
% \texttt{sbi}~\cite{tejero-cantero2020sbi},
% \texttt{scikit-learn}~\cite{JMLR:v12:pedregosa11a},
\texttt{SciPy}~\cite{2020SciPy-NMeth},
and \texttt{tqdm}~\cite{casper_da_costa_luis_2021_5517697} software packages.
%%%%%%%%%%%%%%%

\appendix
\section*{Appendix A: Additional parameter in fit to \Fermi data}
\label{appd:fermi-post-extra}

\begin{figure*}[ht]
    \centering
    \includegraphics[width=0.8\linewidth]{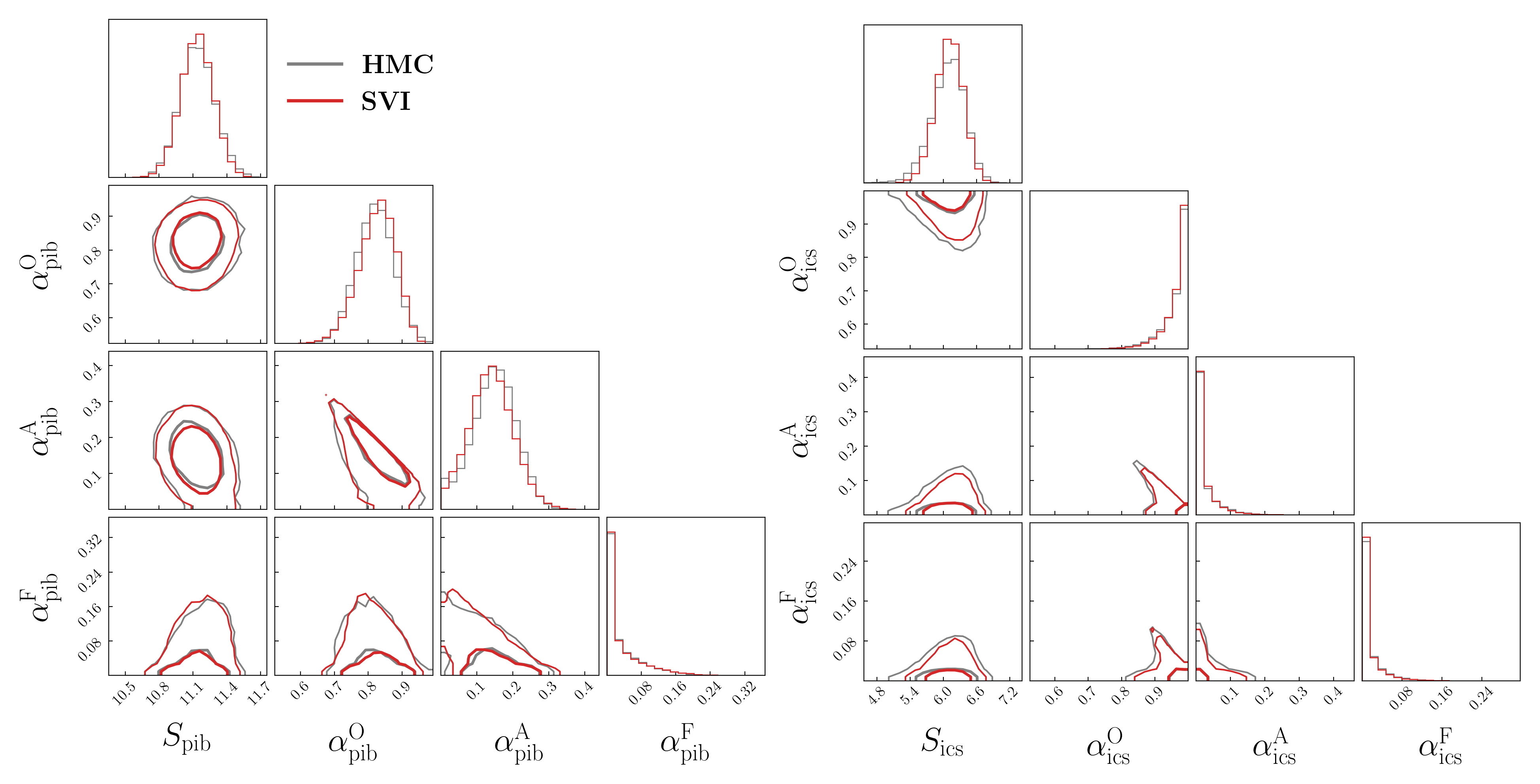}
    \caption{{\textit{Fermi} posterior on gas correlated and ICS mixtures.} Same as Fig.~\ref{fig:fermi-post} but for: (\textit{left}) normalization for $\pi_0$+bremsstrahlung (pib) and mixture fractions, and (\textit{right}) normalization for ICS and mixture fractions. See Subsec.~\ref{subsec:dataset-templates} for details on the templates.}
    \label{fig:fermipibics-post}
\end{figure*}

\begin{figure*}[ht]
    \centering
    \includegraphics[width=\linewidth]{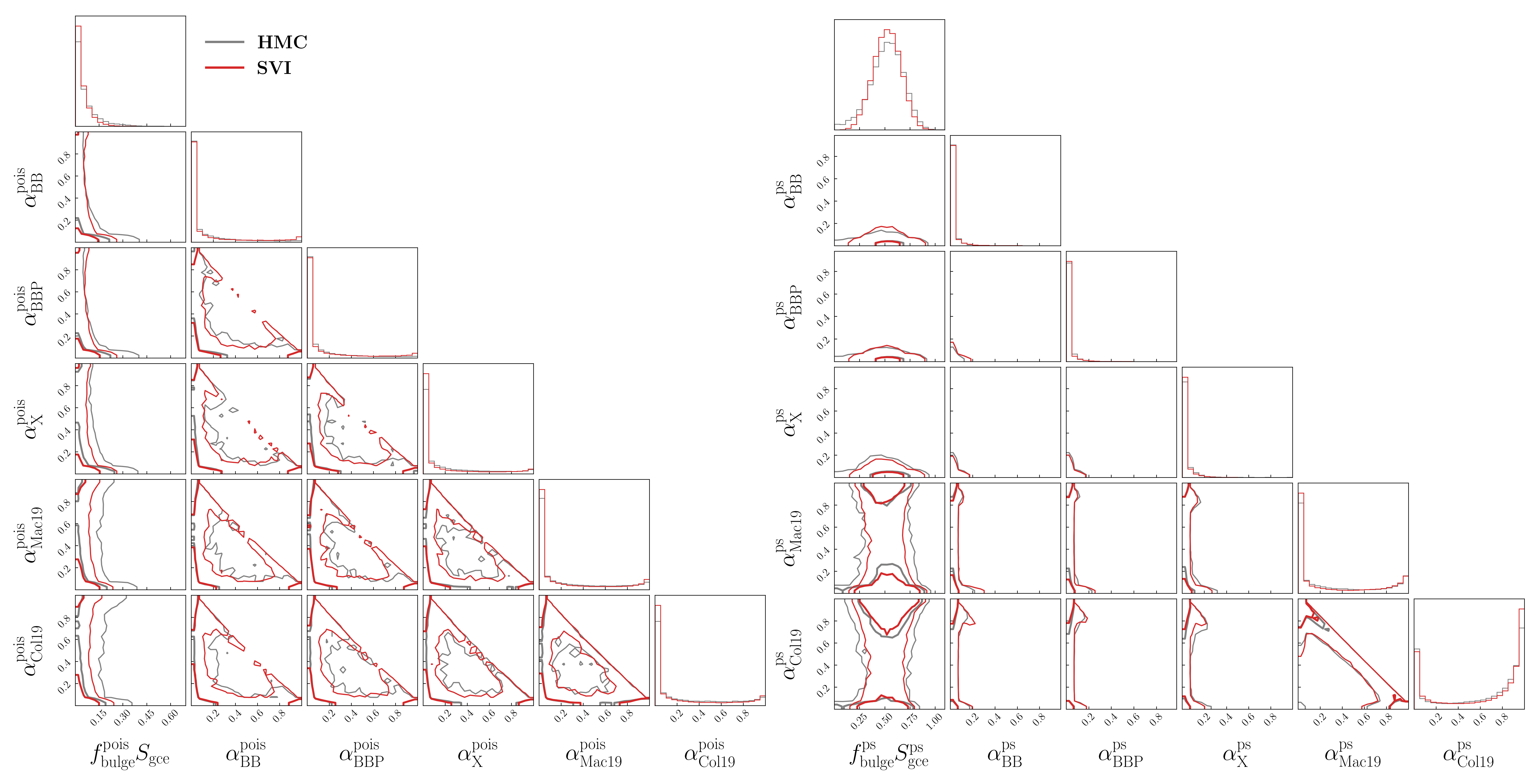}
    \caption{{\textit{Fermi} posterior on diffuse and PS bulge mixtures.} Same as Fig.~\ref{fig:fermi-post} but for: (\textit{left}) normalization of diffuse bulge $f^\text{pois}_\text{bulge}S_\text{gce}$ and mixture fractions, and (\textit{right}) normalization for PS bulge $f^\text{ps}_\text{bulge}S^\text{ps}_\text{gce}$ and mixture fractions. The posterior on $\vec\alpha_\text{pois}$ is largely driven by the Dirichlet prior due to a lack of constraining power, given the small reconstructed normalization for the bulge component. See Subsec.~\ref{subsec:dataset-templates} for details on the templates.}
    \label{fig:fermibulge-post}
\end{figure*}

In this section we augment the plots in the main text with the posterior distributions for additional selected parameters in Fig.~\ref{fig:fermipibics-post} and Fig.~\ref{fig:fermibulge-post}.

\section*{Appendix B: Restricted prior in SVI marginal calibration}
\label{appd:restricted-prior}

The restricted prior, in comparison to the fiducial prior, used in the SVI marginal calibration tests is shown in Tab.~\ref{tab:prior-restricted}. Compared to the fiducial, shown in Tab.~\ref{tab:prior}, the upper limits of $S_\text{pib}$, $S_\text{ics}$, $S_\text{bub}$, $S_\text{psc}$, $S_\text{gce}$, $S^\text{ps}_\text{dsk}$, and $S^\text{ps}_\text{gce}$ have been reduced, while the lower limits and priors on other variables remain unchanged.

\TabPriorRestricted

\begin{figure}[ht]
    \centering
    \includegraphics[width=\linewidth]{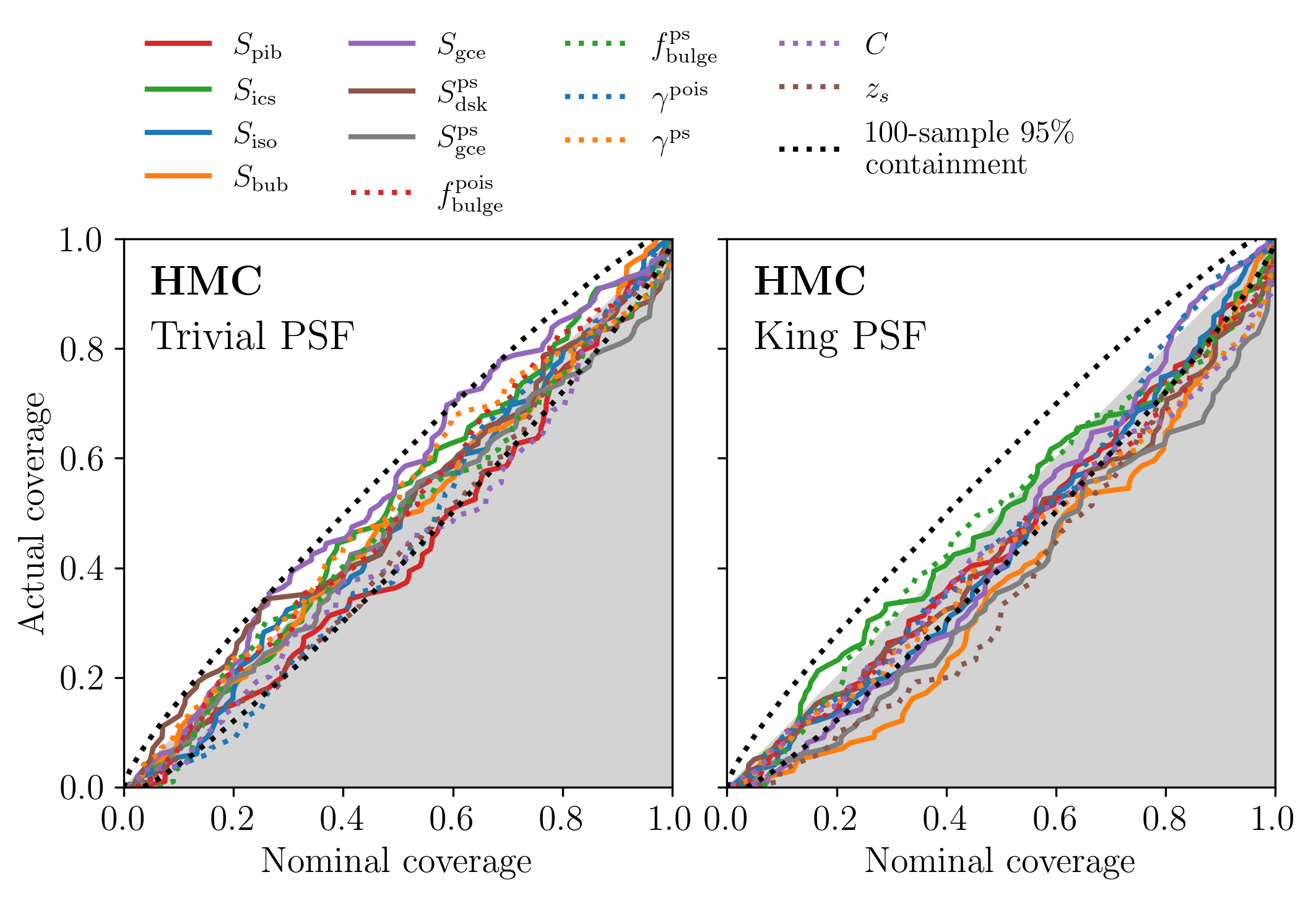}
    \caption{Marginal coverage curves for HMC using the restricted prior, shown as a cross-check that the restricted prior is a fair surrogate.}
    \label{fig:marginal-coverage-hmc-restricted}
\end{figure}

In Fig.~\ref{fig:marginal-coverage-hmc-restricted}, we show that HMC remains well-calibrated in a marginal coverage test with the restricted prior, as a cross check.

\begin{figure*}[ht!]
    \centering
    \includegraphics[width=\linewidth]{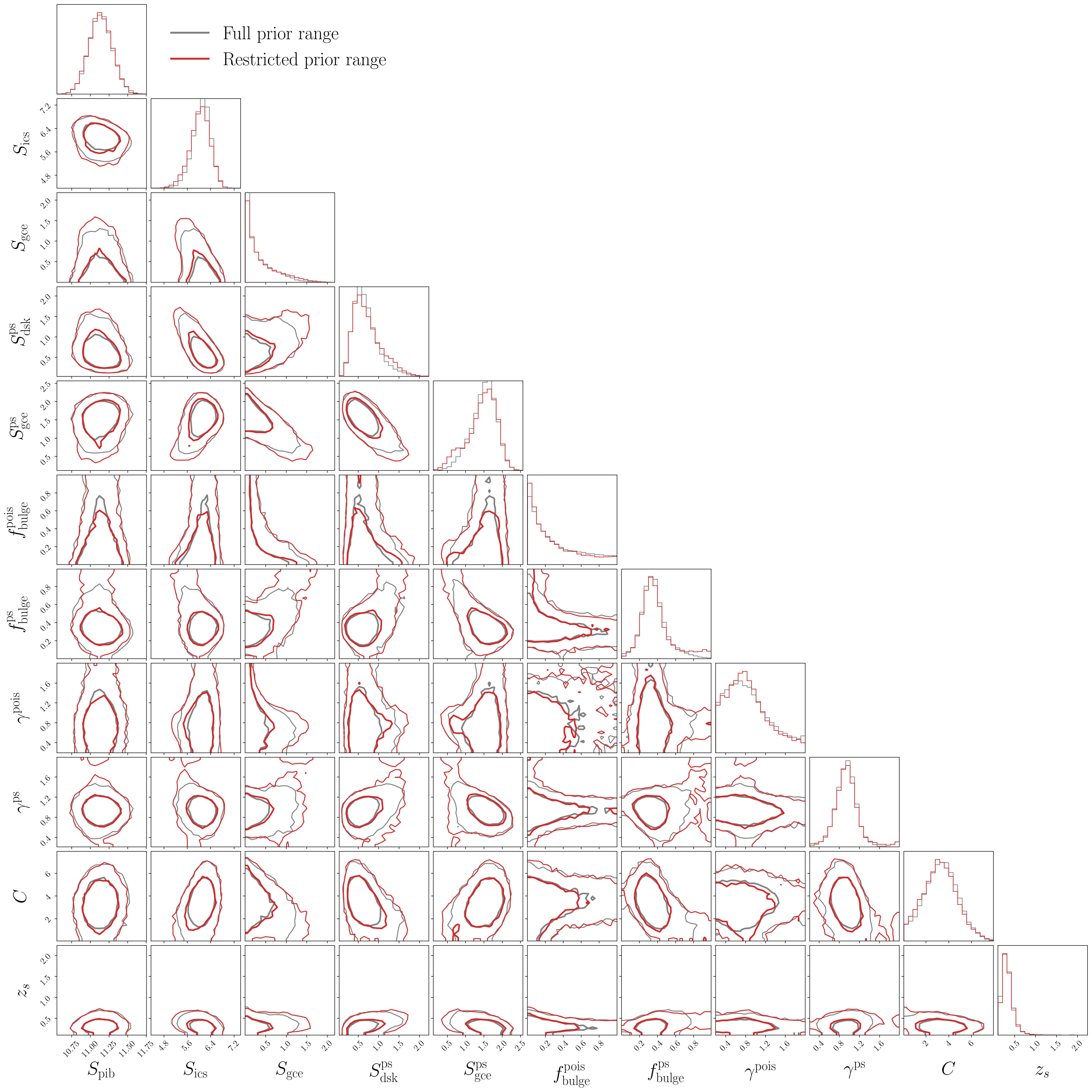}
    \caption{Posteriors from fitting to \Fermi data with the full prior (black) versus the restricted prior used for the SVI calibration test (red), for select parameters. The two are in close agreement, confirming that the restricted prior range is sufficient.}
    \label{fig:fermi-post-restricted}
\end{figure*}

In Fig.~\ref{fig:fermi-post-restricted}, we show that fitting to \Fermi data with the full prior and the restricted prior yield very similar posteriors.

\clearpage

\bibliographystyle{apsrev4-1}
\bibliography{fermi-prob}

\end{document}